\documentclass[11pt,oneside,letterpaper]{article}
\usepackage{amssymb}
\usepackage{amsmath}
\usepackage[dvips]{graphicx}
\usepackage{setspace}
\usepackage{amsfonts}
\usepackage{fancyhdr}
\usepackage{xcolor}
\usepackage{graphicx}
\usepackage{rotating}
\usepackage{comment}
\usepackage{color}
\usepackage{cite}
\usepackage{braket}

\usepackage{tikz}

\usepackage{subfigure}
\usetikzlibrary{decorations.pathmorphing}

\definecolor{darkgreen}{rgb}{0,0.5,0}
\definecolor{darkblue}{rgb}{0,0,0.6}
\definecolor{purple}{rgb}{0.4,.2,0.7}
\definecolor{black}{rgb}{.2,.2,.2}

\definecolor{tblue}{RGB}{74,144,226} 
\definecolor{tred}{RGB}{208,2,27} 
\definecolor{torange}{RGB}{245,166,35} 
\definecolor{darkgreen}{RGB}{126,211,33} 
\definecolor{tblue}{RGB}{74,144,226}
\definecolor{tpurple}{RGB}{189,16,224} 
\definecolor{darkgreen}{RGB}{65,117,5}

\usepackage[colorlinks=true,
citecolor=darkgreen,
linkcolor=black,
urlcolor=purple]{hyperref}

\usepackage{pdfsync}

\makeatletter
\newcommand*{\defeq}{\mathrel{\rlap{%
                     \raisebox{0.3ex}{$\m@th\cdot$}}%
                     \raisebox{-0.3ex}{$\m@th\cdot$}}%
                     =} 
\makeatother

\newcommand{\la}{\langle}

\newcommand{\bea}{\begin{eqnarray}}
\newcommand{\eea}{\end{eqnarray}}

\def\half{{\textstyle{\frac{1}{2}}}}

\let\a=\alpha             \let\n=\nu  \let\r=v        \let\G=\Gamma \let\De=\Delta         
\let\la=\label

\def\be{\begin{equation}}
\def\ee{\end{equation}}
\def\ba{\begin{array}}
\def\ea{\end{array}}

\def\ba#1\ea{\begin{align}#1\end{align}}
\def\bs#1\es{\begin{split}#1\end{split}}

\allowdisplaybreaks  % allow page breaks in displayed eqs

\numberwithin{equation}{section}
\def\nref#1{(\ref{#1})}
\def \la {\label}   % Had to remove your definition for \la 
\def \be {\begin{equation}}
\def \ee {\end{equation}}
\def \half {{1\over 2}}	
\def \JM#1 {{\color{red}  JM: #1 }}
\def \MG#1 {{\color{blue}  MG: #1 }}

\begin{document}
\onehalfspacing

\begin{center}

~
\vskip5mm

{\LARGE  {Proper time to the black hole singularity from thermal one-point functions}  \\
}

\vskip10mm

 Matan Grinberg$^{1,2}$, \ \ Juan Maldacena$^{3}$ 

\vskip15mm
{\it $^{1}$ Department of Physics, Princeton University, Princeton, New Jersey, USA } \\
\vskip5mm

{\it $^{2}$ Department of Applied Mathematics and Theoretical Physics,\\
University of Cambridge, Cambridge, UK} \\
\vskip5mm
 
{\it $^{3}$ Institute for Advanced Study, Princeton, New Jersey, USA } \\
\vskip5mm

\vskip5mm

\end{center}

\vspace{4mm}

\begin{abstract}
\noindent
 
We argue that the proper time from the event horizon to the black hole singularity can be extracted from the thermal expectation values of certain operators outside the horizon. This works for fields which couple to higher-curvature terms, so that they can decay into two gravitons. To extract this proper time, it is necessary to vary the mass of the field.

 \end{abstract}
%\vspace{.2in}
%\vspace{.3in}

\pagebreak
\pagestyle{plain}

\setcounter{tocdepth}{2}
{}
\vfill
\tableofcontents

\newpage

%%%%%%%%%%%%%%%%%%%%%%%%%%%%%%%%%%%%%%%%%%%%%%%%

\section{Introduction}
 
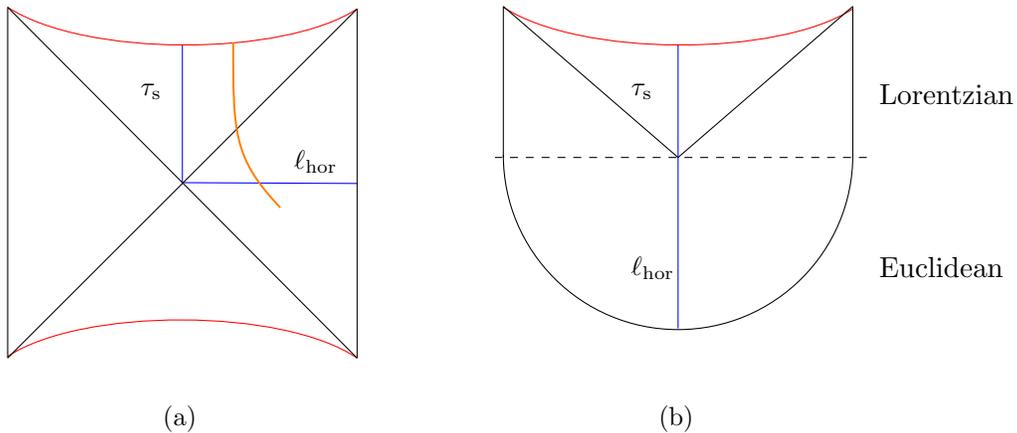
\begin{figure}[h]
\begin{center}
\begin{tikzpicture}[x=0.75pt,y=0.75pt,yscale=-1.25,xscale=1.25]
\draw [color=blue  ,draw opacity=1 ]   (328.85,25.67) -- (328.85,81.5) ;
\draw [color=blue  ,draw opacity=1 ]   (328.85,81.17) -- (399.35,81.5) ;
\draw [color=red  ,draw opacity=1 ]   (258.35,10.33) .. controls (289.35,30.56) and (369.35,30.56) .. (399.35,11) ;
\draw [color=red  ,draw opacity=1 ]   (258.35,151.33) .. controls (288.35,131.56) and (369.35,131.56) .. (399.35,152) ; 
\draw (373,67) node [anchor=north west][inner sep=0.75pt]  [font=\small]  {${\displaystyle \ell _{\rm hor}}$};
\draw (311,40.4) node [anchor=north west][inner sep=0.75pt]  [font=\small]  {$\tau _{\rm s}$};
\draw    (399.35,11) -- (399.35,152) ;
\draw    (258.35,10.33) -- (258.35,151.89) ;
\draw    (399.35,11) -- (258.35,152) ;
\draw    (258.35,10.33) -- (399.35,152) ;
\draw [color=orange  ,draw opacity=1 ][line width=0.75]    (349.35,24.56) .. controls (349.35,61.22) and (349.35,71.22) .. (368.35,91.22) ;
\draw (320,170) node [anchor=north west][inner sep=0.75pt]  [font=\small]  {(a)};
%
%%%%%%%%% Euclidean-Lorentzian Mixed
%
\draw [color=blue  ,draw opacity=1 ]   (528.85,25.67) -- (528.85,140) ;
\draw [color=red  ,draw opacity=1 ]   (458.35,10.33) .. controls (489.35,30.56) and (569.35,30.56) .. (599.35,11) ;
%\draw (573,67) node [anchor=north west][inner sep=0.75pt]  [font=\small]  {${\displaystyle \ell _{\rm hor}}$};
%\draw (511,40.4) node [anchor=north west][inner sep=0.75pt]  [font=\small]  {$\tau _{\rm s}$};
\draw    (599.35,10) -- (599.35,71) ;
\draw    (458.35,10) -- (458.35,71) ;
\draw    (599.35,10) -- (528.85,71) ;
\draw    (458.35,10) -- (528.85,71) ;
\draw[dashed]    (605,71) -- (455,71) ;
%\draw [color=black  ,draw opacity=1 ]   (458.35,71) .. controls (459.35,180.56) and (599.35,180.56) .. (599.35,71) ;
\draw (509,110) node [anchor=north west][inner sep=0.75pt]  [font=\small]  {${\displaystyle \ell _{\rm hor}}$};
\draw (509,40.4) node [anchor=north west][inner sep=0.75pt]  [font=\small]  {$\tau _{\rm s}$};
\draw (599.35,70) arc (0:180:70.5);
\draw (520,170) node [anchor=north west][inner sep=0.75pt]  [font=\small]  {(b)};
\draw (609,110) node [anchor=north west][inner sep=0.75pt]  {Euclidean};
\draw (609,40.4) node [anchor=north west][inner sep=0.75pt] {Lorentzian};
\end{tikzpicture}
\caption{(a) We define $\tau_s$ to be the time between the bifurcation surface and the singularity. $\ell_{\rm hor}$  is the (renormalized) distance from the horizon and the boundary. The time from the horizon to the singularity along any timelike curve, such as the orange curve is smaller than $\tau_s$.     (b) The geodesic relevant for the one point function computation, plotted in the mixed Lorentzian/Euclidean geometry. (We thank Gautam Mandal for suggesting this representation.)}
\label{TimeToSingularity}
\end{center}
\end{figure}

If you are going to fall into a Schwarzschild black hole, it would be helpful to know how long you can live inside. Your lifetime inside is shorter than (or equal to) the time  between the bifurcation surface and the singularity.  See figure \ref{TimeToSingularity}. 
Given that this is an interesting property of a black hole, we would like to be able to extract it by computing properties of correlation functions outside the black hole.

Under some reasonable assumptions,  the simplest correlation function --- the one-point function of a massive field --- contains this information. In particular, one needs to examine the dependence of this expectation value on the mass of the field. 
We argue that the time to the singularity, $\tau_{\rm s}$, is contained in its exponential large-mass behavior
\be 
\langle O \rangle \sim  
({\rm powers~of~}m) \times 
\exp\left[ - i m \tau_{\rm s} - m \ell_{\rm hor} \right]
~,~~~~{\rm for}~~~~
{\rm Im}(m) <0 \, ,
\la{MainEq}
\ee 
where we have assumed that $m$ has a negative imaginary part. In other words, we can say that the time to the singularity arises from a ``phase'' in the one-point function. Of course, the one-point function is real for real $m$, but it develops this ``phase'' for complex $m$\footnote{The word ``phase'' is in quotation marks because, for complex $m$, the term involving $i m \tau_s$ is not a pure phase. It is just the term with an extra $i$ in the exponential.}. This expression requires some assumptions about the coupling of the massive field to gravity, which we specify below. 
 
 Let us first give a quick rationale for this formula and we will make   more precise statements later in the paper. 
 
A minimally coupled field has a quadratic action which leads to a vanishing one-point function. However, a non-zero value could result if higher-derivative corrections to the action, such as a coupling between the field and the squared Weyl tensor, are included. Physically, this means that the particle in question  can decay into two gravitons. This assumption is true if the initial field is a generic massive string mode in string theory.  On a black hole background, this gravitational coupling leads to  a source term for the field and therefore, a one-point function \cite{Myers:2016wsu}\footnote{In three bulk dimensions,  thermal one-point functions arise due to particles wrapping the horizon \cite{Kraus:2016nwo}. }. We are treating the field as a probe of the background, ignoring its backreaction. 
In the large-mass approximation, correlation functions of the field can be approximated in terms of geodesics. The geodesic starts at the insertion point of the operator. The other end is integrated over spacetime, weighed by the background value of the squared Weyl tensor. In a saddle-point approximation, we should balance the ``force'' from the geodesic with that exerted by the spatial variation of the Weyl tensor. Because of the large mass, the geodesic contribution dominates everywhere except very close to the singularity. For this reason, the saddle point is at a (complex) radial position very near the singularity. Therefore, the saddle-point approximation gives us the time to the singularity as in \nref{MainEq}. See figure \ref{TimeToSingularity}. The real part in the exponent involves the distance from the operator insertion to the horizon. 
 
There are some further details and qualifications that we will spell out later in the paper. In the context of a simple example of AdS/CFT, such as the case of ${\cal N}=4$ four-dimensional supersymmetric Yang-Mills \cite{Maldacena:1997re,Witten:1998qj,Gubser:1998bc}, we are considering one-point functions on a black hole background  to leading order in the large-$N$ approximation. The massive field can be a massive string state in the bulk with mass on the order of the string scale. The mass can be varied by varying the t' Hooft coupling of the gauge theory, since $m \propto \lambda^{1/4}$ \cite{Gubser:1998bc,Witten:1998qj}. We can also give it an imaginary part by taking $\lambda $ complex, in which case \nref{MainEq} holds.  
 
Our discussion is in the spirit of \cite{Fidkowski:2003nf}, though the analytic continuation we use   looks a bit simpler. The information we get is also simpler. We only claim that it gives us the time to the singularity. On the other hand, the procedure in \cite{Fidkowski:2003nf} gives  a more direct signal from the singularity. 
 
The rest of the paper is organized as follows. In section two, we explain how higher-derivative corrections give rise to thermal one-point functions. In section three, we discuss how to compute the one-point functions for large mass by using a geodesic approximation. In section four, we discuss in detail the example of a black brane. In section five, we explore various geodesics that can contribute for more general black holes. In section six, we discuss some aspects of black holes with inner horizons. Finally, we present some conclusions.

  %Interesting aspects of thermal one-point functions include... \JM{This the ``thermal inversion formula'' \cite{Iliesiu:2018fao}.   OPE and thermal 1pt function \cite{Gerchkovitz:2016gxx} (also ElShowk and Papapadodimas).  Thermal blocks: \cite{Gobeil:2018fzy} } 
  
\section{One-point functions from higher-derivative corrections } 

We consider the Lagrangian 
\be \la{LagraSta}
 S= { 1 \over 16 \pi G_{\rm N} } \int \left[ \half (\nabla \varphi)^2 + \half m^2 \varphi^2  + \alpha \varphi W^2 \right] \,,
 \ee 
containing a single massive field and the simplest higher-derivative coupling to the gravitational field. Here $W^2 = W_{\mu \nu \delta \sigma} W^{\mu \nu \delta \sigma}$ is the square of the Weyl tensor\footnote{Couplings to the Ricci scalar or Ricci tensor can be removed by field redefinitions.}.   Here we consider a scalar field, but one can write similar couplings for higher-spin fields. 
 The coupling $\alpha$ is  expected to be small,   $\alpha \propto  \alpha' \propto { 1 \over \sqrt{\lambda} }$. \footnote{Causality based bounds on $\alpha$ were discussed in \cite{Cordova:2017zej,Meltzer:2017rtf}.} Note that \nref{LagraSta} is a coupling that appears in the classical theory at leading order in the $G_{\rm N}$ expansion. 

Let us consider this theory in AdS$_{d+1}$. In AdS, $W^2=0$ and the one-point function is also zero, as generically required by conformal symmetry.  On the other hand, for a black hole the Weyl tensor is nonzero, and this nonzero value  sources the field $\varphi$. In Euclidean space, we can write the one-point function as 
\be \la{IntExp} 
\langle O(0) \rangle \propto \alpha \int_{\rm EBH}d^{d+1} x \, \sqrt{g} \,  G(0;x)\,  W^2 \, ,
\ee 
where $G$ is the boundary-to-bulk propagator for the massive field. The integral is over the Euclidean black hole. This integral is convergent for small enough masses, namely, 
$\Delta < 2 d$. 
Here, $\Delta $ is the scaling dimension, given by \cite{Gubser:1998bc,Witten:1998qj}
\be  \Delta = { d \over 2} + \sqrt{ {d^2 \over4 } + m^2 R^2}
~,~~~~~ \Delta \sim mR
~,~~~~~{\rm for } ~~ mR \gg 1 \,.
\ee  
For $\Delta > 2 d$, the integral \nref{IntExp} diverges.
 This divergence is a common feature of AdS Witten diagrams involving  fields that can decay into lighter fields. In this case, the field $\varphi$ can decay into two gravitons. By analytically continuing in the dimension, we can define finite integrals, as is standard \cite{Freedman:1998tz}. In this case, the resulting function has poles at certain values of $\Delta$. These values are the dimensions of multi-graviton operators that have non-zero vacuum expectation values in the black hole background. One possible sequence of operators corresponds to powers of the stress tensor and lead to poles at $\Delta = n d$, for $n\geq 2$. These poles result from enhanced operator mixing when there is a ``resonance''. For generic operator dimensions, the mixing is suppressed by powers of $1/N$. However, if two dimensions coincide, then we can have mixing at leading order in the large-$N$ expansion. The fact that this mixing is larger than for generic dimensions leads to poles in correlators as function of the dimension.  More precisely, when we think about the regularized version of the operator $O$, it can mix with lower-dimension operators. When we compute the one-point function, we are interested in the one-point function of the operator with the large dimension. In the large-$N$ limit, this is well defined as long as $\Delta$ is not at one of the resonant dimensions. See appendix \ref{OpMix}.

\section{One-point functions from the geodesic approximation} 

For large mass, $m R_{\text{AdS}} \sim \Delta \gg 1$,    we can  use the  geodesic approximation for the propagator in \nref{IntExp}. This amounts to approximating 
\be
G \sim e^{ -m  \ell } \la{Geo}\, ,
\ee
 where $\ell$ is the (renormalized) proper length between the boundary and a bulk point. The prefactor in \nref{Geo} can also be written down, see appendix \ref{Prefactor}.

When we insert this into  \nref{IntExp}, we find that the propagator has a strong dependence on position due to the large exponent in \nref{Geo}. Furthermore,  \nref{Geo}  is strongly peaked near the boundary, which leads to a divergence there --- the same one we mentioned above when $\Delta$ is large. This divergent contribution can be interpreted as arising from the $\varphi$ particle decaying into gravitons near the boundary, which gives us the expectation value of the corresponding multi-trace operator of the stress tensor. Notice that a conceptually similar feature arises when we compute the vacuum AdS three-point functions between $O$ and two stress tensors, using the geodesic approximation. This integral near the boundary gives rise to the poles in the three-point function. In appendix \ref{ThreePoints} we discuss a simple example.  

The interesting contribution to the one-point function comes from a solution where we balance the pull from the propagator \nref{Geo} and the $W^2$ term. 
This can happen only where the $W^2$ term is varying rapidly. This does not happen anywhere in the Euclidean black hole. However, we can analytically continue the integral to the region near the singularity where $W^2$ is diverging and thus,  we can find a balance between the two terms.  In order to continue the geodesic beyond the horizon, we need to pick a branch. We must decide whether to continue it as $\ell = \ell_{\rm hor} + i \tau$ or as $\ell = \ell_{\rm hor} - i \tau$. This is selected by giving an imaginary part to $m$, say $m = m - i \epsilon$. Then one of these continuations results in a decreasing exponential, the one with $\ell_{\rm hor} + i \tau$. This decreasing exponential is what we expect from a saddle-point evaluation and we will later justify it more explicitly in a special case. 
 
At the saddle point, we have the equation 
\be \la{Saddp}
-i m + \partial_{\tau } \log W^2 =0 
~~~~\to ~~~
 i m + { c \over \tau_* -\tau_{\rm s}  } =0
~~~ \to ~~~
\tau_* -\tau_{\rm s}  = - { i c \over m } \, .
\ee 
where $c$ is an order one positive constant. 
%At the saddle point, we have the equation 
%\be \la{Saddp}
%-i m + \partial_{\tau } \log W^2 =0 
%~~~~\to ~~~
%- i m + { 1 \over \tau_* -\tau_{\rm s}  } =0
%~~~ \to ~~~
%\tau_* -\tau_{\rm s}  = - { i \over m } \, .
%\ee 
 We see that for large mass, the saddle point $\tau_*$ is near the singularity at $\tau_{\rm s}$. The displacement away from the singularity is imaginary. This implies that we cannot view this as a point in the Lorentzian black hole. Still, it is close to the singularity in the sense that the leading-order approximation for the integral is given by evaluating 
\be
 \la{GeoGen}
 \langle O \rangle 
 ~~\propto ~~ 
 \sqrt{g(\tau_*)} W^2(\tau_*) e^{ - m l(\tau_*) } 
 ~~\propto  ~~
 \exp\left[ - m \, \ell_{\rm hor} - i m
 \,  \tau_{\rm s} \right]  \times ({\rm powers ~of~} m)\,.
\ee
 The first term comes from evaluating the propagator at the singularity. The deviation away from the singularity in \nref{Saddp} gives a subleading correction. Similarly, the $W^2$ term in \nref{Saddp} only gives powers of $m$. 
We see then, that the small displacement in the imaginary direction in \nref{Saddp} is not important and the final answer involves the time to the singularity. 
 
Let us make some comments:
 \begin{itemize}
 \item
 We are using the fact that we can vary $m$ in order to focus on the $m$-dependence of the correlator. This is appropriate in the case of black holes in string theory, where we can keep the black hole metric fixed and vary the string length, which varies the mass of the fields. 
 \item 
In the particular case of ${\cal N}=4$ supersymmetric Yang-Mills, the change in the mass of the field, or the string length, can be achieved by varying the 't Hooft coupling of the theory.
 \item  
 When we claim that the exponential dependence on $m$ only comes from the propagator, we are assuming that the coupling $\alpha$ in the Lagrangian \nref{LagraSta} does {\it not}  itself have an exponential dependence on $m$. Indeed, in the ${\cal N}=4$ SYM example, it has only a power law dependence on the coupling. Generically in string theory, it is expected to have a power-law dependence on the string length. 
 \item 
 The dependence of the one-point function on the temperature, or the mass of the black hole,  is contained within   
 $\ell_{\rm hor}$. We will see examples below.
 \item 
 We have not shown that the particular saddle point we picked is the dominant one, or that it even contributes.  We will return to this question later.  Depending on the size of the imaginary part of $m$, other saddles can contribute more. 
 \item
Until now, we have discussed the case of a Schwarzschild black hole with its spacelike singularity. We will later discuss black holes with inner horizons. 
 \end{itemize}

\section{Thermal one-point functions for planar black branes}

In this section we consider black branes in various dimensions. In this case, we can do the analytic computation as well as the geodesic analysis. We find a match between these two approaches. 

\subsection{Analytic computation} 
\la{AnBB} 

This computation was done in \cite{Myers:2016wsu}\footnote{ \cite{Myers:2016wsu} did the $d=3$ case, but, as we will see, the hard part is the same for all $d$.} and we now review it. The black brane metric in AdS$_{d+1}$ is 
\be  \la{metric}
ds^2 ={  R^2 \over z^2 } \left[ -f(z) dt^2 + { dz^2\over f(z) } +  d \vec x^2  \right] ~,~~~~~~~f(z) \equiv 1 - { z^d \over z_0^d} ~,~~~~z_0 =  {d \over 4\pi}  \beta\,.
\ee
Since the temperature is the only scale,  the temperature dependence is fixed as
\be  \la{TempDep}
\langle O \rangle \propto z_0^{-\Delta } \propto T^{\Delta } \,.
\ee
We are then left with the problem of fixing the overall coefficient. For this purpose, we can adjust the temperature so that $z_0=1$. 

To construct the propagator, we solve the wave equation with only radial dependence. 
After defining 
\be 
 h \equiv { \Delta \over d } ~,~~~~~~~~w \equiv { z^d \over z_0^d}\,,
 \ee 
 we find equations that are independent of $d$.  We pick two solutions, one regular at infinity and one at the horizon:
\bea
g_{\rm inf}(w)&=& w^h ~_2F_1(h,h,2h;w)\, ,
\\
g_{\rm hor}(w)&=& w^h ~_2F_1(h,h,1;1-w)\, .
\eea 
We then construct the Green's function for the canonically normalized field as
\be \la{propaBB}
G(w,w')=-{1\over R^{d-1} d}{\Gamma(h)^2 \over \Gamma(2h)}\left(g_{\rm inf}(w)g_{\rm hor}(w')\theta(w'-w)
+g_{\rm inf}(w')g_{\rm hor}(w)\theta(w-w')\right)\,.
\ee
Inserting this into \nref{IntExp} and using the fact that, for   
the metric \nref{metric}, the squared Weyl tensor is 
\be 
\label{AdS5Weyl}
W^2 = {d(d-2)(d-1)^2 \over R^4}\left(z\over z_0\right)^{2d} \propto w^2,
\ee we get the final expression for the one-point function:
\bea 
\langle O \rangle &=& 
- C_{\rm N}\sqrt{16 \pi G_{\rm N} \over R^{d-1}}{\a \over  R^2} {(d-2)(d-1)^2 \over   \, d} {\Gamma(h)^2 \over \Gamma(2h)} 
\int_0^1 dw ~w^h~_2F_1( h,h,1; 1-w  ) \cr &=& 
-C_{\rm N}\sqrt{16 \pi G_{\rm N}\over R^{d-1}} {\pi \a \over R^2}  \left( {4 \pi T \over d }\right)^\Delta {(d-2)(d-1)^2 \over  \, d} {\Gamma(h)^2 \over \Gamma(2h)} 
{ h (1-h) \over \sin \pi h }\, .
\la{IntRe}
\eea
In the second line,  we have restored temperature dependence by using \nref{TempDep} and inserting an extra factor of $z_0^{-\Delta}$.
Here, $C_{\rm N}$ is a  normalization coefficient that depends on how we normalize the operator $O$, see  \eqref{CNdef}. Note that $C_{\rm N}$ does not have an exponential dependence on $h$.\footnote{ We can also do this computation more generally for a $\left(W^2\right)^{1+k}$ coupling. It involves, $ 
\int_0^1 dw w^{h+2k}~_2F_1(h,h,1;1-w)
={\Gamma(1+h+2k)\Gamma(2-h+2k) / \Gamma(2+2k)^2}. $
%\be
%\int_0^1 dw w^{h+2k}~_2F_1(h,h,1;1-w)
%={\Gamma(1+h+2k)\Gamma(2-h+2k) \over \Gamma(2+2k)^2}.
%\ee
This correctly reproduces \eqref{IntRe} for $k=0$.} In the final expression   \nref{IntRe},  the only important factor  for us will be the $1/\sin \pi h$.

We see that \nref{IntRe}  has poles at $\Delta = n d$ for $n\geq 2$. The integral expression is convergent only for $\Delta < 2 d$, since the hypergeometric function behaves like $w^{-2 h+1}$ for small $w$.  Here we defined the integral by analytically continuing $h$. The small-$w$ region, which gives rises to the divergences and the poles, corresponds to the region near the boundary of AdS$_{d+1}$. As previously mentioned, we interpret the poles as arising from mixing with operators that are powers of the stress tensor, $T^n$, $n \geq 2$. For black branes, operators involving derivatives of the stress tensor are zero \cite{Iliesiu:2018fao,Gobeil:2018fzy}, implying that expectation values of operators of the schematic form  $T \partial^{2m} T $ vanish\footnote{In this expression, the derivatives are acting on both factors of $T$ in such a way as to yield a conformal primary.}. 

Giving a small negative imaginary part to  $\Delta$ (or equivalently to $h$ or $m$), we can avoid these poles and obtain a large-$|\Delta|$ exponential behavior of the form (for $z_0=1$):
\be \la{IntLa}
\langle O \rangle 
~~\sim~~
e^{-i\pi\left({\Delta \over d }\right)} \, 4^{ -\left({\Delta \over d }\right)}
~~\sim~~ 
e^{-i\pi \left({m R\over d }\right)} 4^{- \left({mR\over d }\right) }
~,~~~~~ {\rm for } ~~ 
{\rm Im}(\Delta) \propto  {\rm Im}(m) < 0
\, .
\ee 
If the imaginary part of $m$ had been positive, we would need to change $i \to -i$ in \nref{IntLa}.  The factor of $4^{-\Delta}$ comes from the gamma functions in \nref{IntRe}. 
 
Notice that \nref{IntRe} is real for real $h$. However, as we give $h$ a small negative imaginary part, one of the exponentials in the sine factor of \nref{IntRe} dominates  and gives rise to the ``phase'' in \nref{IntLa}.

\subsection{Geodesic approximation} 
\la{GeoAp} 

The candidate saddle-point approximation \nref{GeoGen} involves the integral 
\be 
\ell = R\int_0^\infty { dz \over z \sqrt{ 1 - z^d } }   = { R \over d } \int_0^\infty  {dw \over w \sqrt{ 1 -w } }
~,~~~~~{\rm for}~~
 z_0 =1.
\ee 
This integral diverges at small $w$, which is the region near the boundary. In addition, we must decide how to go around the branch cut that starts at the horizon, $w=1$. The singularity is at $w =\infty$. The small-$w$ divergence can be regularized in the same way as the divergence of the two-point function in empty AdS, see appendix \ref{GravNor}.  
This gives the renormalized length to the singularity as
\be \la{RenLe}
\ell = \ell_{\rm hor} - i \tau_{\rm s} =   { R \over d } \lim_{w_{\rm c} \to 0}\left[     \int_{w_{\rm c}}^\infty  { d w \over w \sqrt{ 1 -w } }  + \log w_{\rm c}     \right] = { R \over d } \left[ -  i \pi+ \log (4) \right] .
\ee  
To compute the imaginary part, we had to pick a path around the cut starting at $w=1$ in the first integral. We could have chosen either sign and we discuss this choice in the next subsection. 

Setting the one-point function to $\langle O \rangle \sim e^{ -  m  \ell} $, we reproduce \nref{IntLa}. We see that the factor of $4^{ -\Delta/d}$ comes from the renormalized length up to the horizon. When $z_0\not =1$, this also reproduces the temperature dependence.   
 
\subsection{A more detailed saddle-point analysis } 
\la{SaddleAn}
 Here we sketch a more systematic saddle-point analysis for the integral 
 \nref{IntRe}.
It is convenient to choose a variable $\rho$ defined by
\be \la{Choice}
z^d =w = { 1 \over ( \cosh { \rho \over 2 } )^2 } \, ,
\ee
which is such that we can interpret $\rho$ as proportional to the proper distance from the horizon\footnote{The actual proper distance is ${ R \over d }\rho$.}.

The first step is to define the integral for large $h$. The problem here is the divergence near $w=0$, where the integrand behaves like
\be \la{TwoDiv}
 \int d w ( A w ^{ 1-h} + B w^h ) \sim \int d\rho \left( A e^{(h-2) \rho} + B e^{ -(h-1) \rho } \right) ~,~~~~~{\rm for}~~
 w\ll 1
~,~~~~~{\rm or}~~
\rho \gg 1\,.
\ee
For small $w$ and large positive $\rho$, we have that $w \propto  e^{ - \rho} $. This is the region near the AdS boundary. 
If we make $h$ complex, then it is possible to make the integral convergent by tilting the integration contour into the imaginary direction.  It is possible to tilt it in such a way that both terms are convergent. This defines a convergent integral, which is the integral that we would like to approximate using the saddle-point method. 

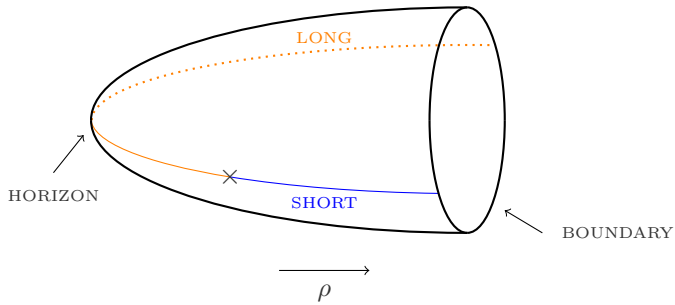
\begin{figure}[h]
\centering
\begin{tikzpicture}
\draw[orange, dotted, thick] (-5.49,0) arc (180:86:5cm and 1cm);
\draw[orange] (-5.5,.02) arc (180:231:5cm and 1cm);
\draw[blue] (-.875,-.975) arc (-94.2:-129:5cm and 1cm);
\filldraw[blue] (-2.4,-1.1) node[] {\tiny SHORT};
\filldraw[orange] (-2.4,1.1) node[] {\tiny LONG};
\filldraw[black] (-2.4,-2.3) node[] {\small $\rho$};
\draw[->] (-3,-2) -- (-1.8,-2);
\filldraw[black] (-6,-1) node[] {\tiny HORIZON};
\draw[->] (-6,-.7) -- (-5.6,-.2);
\filldraw[black] (1.5,-1.5) node[] {\tiny BOUNDARY};
\draw[->] (.5,-1.5) -- (0,-1.2);
\draw[thick] (0,0) arc (0:90:.5cm and 1.5cm);
\draw[thick] (0,0) arc (0:-90:.5cm and 1.5cm);
\draw[thick] (-1,0) arc (180:90:.5cm and 1.5cm);
\draw[thick] (-1,0) arc (180:270:.5cm and 1.5cm);
\draw[thick] (-.5,-1.5) arc (270:180:5cm and 1.5cm);
\draw[thick] (-.5,1.5) arc (90:180:5cm and 1.5cm);
\filldraw[black] (-3.65,-.76) node[] {$\times$};
\end{tikzpicture}
\caption{Euclidean cigar with two geodesics: long (orange) and short (blue).}
\label{Geodesics}
\end{figure}

We fix a bulk point, and look for a geodesic that goes from  this point to the point on the boundary where the operator is inserted. We are supposed to integrate over the angular direction of the bulk point. This integral is the same as integrating over the position of the insertion of the boundary operator. So we fix an arbitrary position of the angular coordinate for the bulk point, but we allow the geodesic to end at any value of the euclidean time direction on the boundary. The point where its ends will be chosen by minimizing the length of the geodesic.   There are two geodesics that go from a given bulk  point to the boundary. One goes straight to the boundary, the other goes to the horizon (the tip of the cigar) and then to the boundary. We call them the ``short'' and the ``long'' geodesics, respectively, see figure \ref{Geodesics}. We need to sum over the contributions of these two geodesics.
% One might worry that they end at different points on the boundary. However, as we integrate over the angular position of the Euclidean time coordinate of the bulk point, each geodesic will intersect a given boundary point just once. 
These long and short geodesics are responsible for the two terms we have in \nref{TwoDiv}. The second term is the long geodesic contribution, which is convergent  at large $\rho$. The first term is the short geodesic contribution, which diverges on the original contour.

%We fix a bulk point, and look for a geodesic that goes from the boundary to that point.
%There are two geodesics that go from that point to the boundary. One goes straight to the boundary, the other goes to the horizon (the tip of the cigar) and then to the boundary. We call them the ``short'' and the ``long'' geodesics, respectively, see figure \ref{Geodesics}. We need to sum over the contributions of these two geodesics. One might worry that they end at different points on the boundary. However, as we integrate over the angular position of the Euclidean time coordinate of the bulk point, each geodesic will intersect a given boundary point just once. 
%These long and short geodesics are responsible for the two terms we have in \nref{TwoDiv}. The second term is the long geodesic contribution, which is convergent  at large $\rho$. The first term is the short geodesic contribution, which diverges on the original contour. 

After we choose the tilted contour indicated above, we can rotate the contour differently for the short and long geodesic contributions. For the long geodesic, we simply bring it to the original position, at $\rho \in [0, \infty]$. For the short geodesic, we must approach the continuation more carefully.

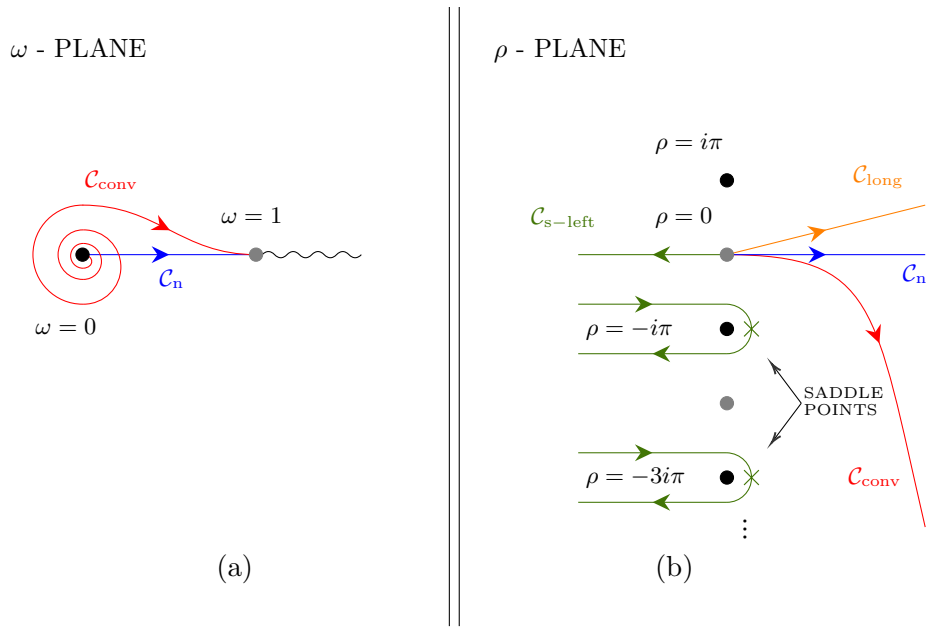
\begin{figure}[h]
\begin{center}
\begin{tikzpicture}[x=0.75pt,y=0.75pt,yscale=-1.25,xscale=1.25]
\draw    (332,0) -- (332,250)(328,0) -- (328,250) ;
\draw [color=darkgreen  ,draw opacity=1 ]   (380,100) -- (440,100) ;
\draw [shift={(410,100)}, rotate = 0] [fill=darkgreen  ,fill opacity=1 ][line width=0.08]  [draw opacity=0] (7.14,-3.43) -- (0,0) -- (7.14,3.43) -- (4.74,0) -- cycle    ;
\draw [color=darkgreen  ,draw opacity=1 ]   (440,120) -- (380,120) ;
\draw [color=darkgreen  ,draw opacity=1 ]   (380,140) -- (440,140) ;
\draw [color=darkgreen  ,draw opacity=1 ]   (440,180) -- (380,180) ;
\draw [color=darkgreen  ,draw opacity=1 ]   (380,200) -- (440,200) ;
\draw [shift={(410,120)}, rotate = 180] [fill=darkgreen  ,fill opacity=1 ][line width=0.08]  [draw opacity=0] (7.14,-3.43) -- (0,0) -- (7.14,3.43) -- (4.74,0) -- cycle    ;
\draw [shift={(410,140)}, rotate = 0] [fill=darkgreen  ,fill opacity=1 ][line width=0.08]  [draw opacity=0] (7.14,-3.43) -- (0,0) -- (7.14,3.43) -- (4.74,0) -- cycle    ;
\draw [shift={(410,180)}, rotate = 180] [fill=darkgreen  ,fill opacity=1 ][line width=0.08]  [draw opacity=0] (7.14,-3.43) -- (0,0) -- (7.14,3.43) -- (4.74,0) -- cycle    ;
\draw [shift={(410,200)}, rotate = 0] [fill=darkgreen  ,fill opacity=1 ][line width=0.08]  [draw opacity=0] (7.14,-3.43) -- (0,0) -- (7.14,3.43) -- (4.74,0) -- cycle    ;
\draw  [color=darkgreen  ,draw opacity=1 ][fill=darkgreen  ,fill opacity=1 ] (446.9,127) -- (452.9,133)(452.9,127) -- (446.9,133) ;
\draw  [color=darkgreen  ,draw opacity=1 ][fill=darkgreen  ,fill opacity=1 ] (446.9,187) -- (452.9,193)(452.9,187) -- (446.9,193) ;
\draw  [color=darkgreen  ,draw opacity=1 ] (439.98,120) .. controls (445.59,120.21) 
and (449.98,124.67) .. (449.94,130.14) .. controls (449.9,135.6) and (445.46,140.01) .. (439.98,140) ;
\draw  [color=darkgreen  ,draw opacity=1 ] (439.98,180) .. controls (445.59,180.21) and (449.98,184.67) .. (449.94,190.14) .. controls (449.9,195.6) and (445.46,200.01) .. (439.98,200) ;
\draw [color=orange  ,draw opacity=1 ]   (440,100) -- (520,80) ;
\draw [shift={(480,90)}, rotate = 525.96] [fill=orange  ,fill opacity=1 ][line width=0.08]  [draw opacity=0] (7.14,-3.43) -- (0,0) -- (7.14,3.43) -- (4.74,0) -- cycle    ;
\draw [color=red ,draw opacity=1 ]   (440,100) .. controls (500.36,100.19) and (499.36,120.19) .. (520,210) ;
\draw [shift={(501.53,135.78)}, rotate = 248.58] [fill=red  ,fill opacity=1 ][line width=0.08]  [draw opacity=0] (7.14,-3.43) -- (0,0) -- (7.14,3.43) -- (4.74,0) -- cycle ;
\draw [color=blue  ,draw opacity=1 ]   (440,100) -- (520,100) ;
\draw [shift={(480,100)}, rotate = 180] [fill=blue  ,fill opacity=1 ][line width=0.08]  [draw opacity=0] (7.14,-3.43) -- (0,0) -- (7.14,3.43) -- (4.74,0) -- cycle    ;
\draw    (470,160) -- (458,176) ;
\draw [shift={(458,176)}, rotate = 307] [color=black  ][line width=0.75]    (4.37,-1.32) .. controls (2.78,-0.56) and (1.32,-0.12) .. (0,0) .. controls (1.32,0.12) and (2.78,0.56) .. (4.37,1.32)   ;
\draw    (470,160) -- (458,144) ;
\draw [shift={(458,144)}, rotate = 413] [color=black  ][line width=0.75]    (4.37,-1.32) .. controls (2.78,-0.56) and (1.32,-0.12) .. (0,0) .. controls (1.32,0.12) and (2.78,0.56) .. (4.37,1.32)   ;
\draw [color=blue  ,draw opacity=1 ]   (180,100) -- (250,100) ;
\draw [shift={(215,100)}, rotate = 180] [fill=blue  ,fill opacity=1 ][line width=0.08]  [draw opacity=0] (7.14,-3.43) -- (0,0) -- (7.14,3.43) -- (4.74,0) -- cycle    ;
\draw    (252.5,100) .. controls (254.17,98.33) and (255.83,98.33) .. (257.5,100) .. controls (259.17,101.67) and (260.83,101.67) .. (262.5,100) .. controls (264.17,98.33) and (265.83,98.33) .. (267.5,100) .. controls (269.17,101.67) and (270.83,101.67) .. (272.5,100) .. controls (274.17,98.33) and (275.83,98.33) .. (277.5,100) .. controls (279.17,101.67) and (280.83,101.67) .. (282.5,100) .. controls (284.17,98.33) and (285.83,98.33) .. (287.5,100) .. controls (289.17,101.67) and (290.83,101.67) .. (292.5,100) -- (292.5,100) ;
\draw [shift={(215.3,90.15)}, rotate = 206.49] [fill=red,fill opacity=1 ][line width=0.08]  [draw opacity=0] (7.14,-3.43) -- (0,0) -- (7.14,3.43) -- (4.74,0) -- cycle    ;
\draw [color=red  ,draw opacity=1 ]   (180,80) .. controls (210.5,80) and (219.5,100) .. (250,100) ;
\draw  [color=red  ,draw opacity=1 ] (180.24,120) .. controls (180.16,120) and (180.08,120) .. (180,120) .. controls (168.95,120) and (160,111.05) .. (160,100) .. controls (160,88.95) and (168.95,80) .. (180,80) .. controls (180.11,80) and (180.21,80) .. (180.32,80) ;
\draw  [color=red  ,draw opacity=1 ] (180.3,90) .. controls (180.36,90) and (180.42,90) .. (180.48,90) .. controls (188.76,90) and (195.48,96.72) .. (195.48,105) .. controls (195.48,113.29) and (188.76,120) .. (180.48,120) .. controls (180.4,120) and (180.32,120) .. (180.24,120) ;
\draw  [color=red  ,draw opacity=1 ] (180.26,110) .. controls (180.22,110) and (180.18,110) .. (180.14,110) .. controls (174.62,110) and (170.14,105.52) .. (170.14,100) .. controls (170.14,94.48) and (174.62,90) .. (180.14,90) .. controls (180.19,90) and (180.25,90) .. (180.3,90) ;
\draw  [color=red  ,draw opacity=1 ] (180.29,95.48) .. controls (180.32,95.48) and (180.35,95.48) .. (180.37,95.48) .. controls (184.39,95.48) and (187.64,98.73) .. (187.64,102.74) .. controls (187.64,106.75) and (184.39,110) .. (180.37,110) .. controls (180.34,110) and (180.3,110) .. (180.26,110) ;
\draw  [color=red  ,draw opacity=1 ] (180.27,105.48) .. controls (180.25,105.48) and (180.23,105.48) .. (180.21,105.48) .. controls (177.45,105.48) and (175.21,103.24) .. (175.21,100.48) .. controls (175.21,97.72) and (177.45,95.48) .. (180.21,95.48) .. controls (180.24,95.48) and (180.26,95.48) .. (180.29,95.48) ;
\draw [color=red ,draw opacity=1 ]   (180.27,105.48) .. controls (184.9,105.24) and (184.9,100.24) .. (180.14,100) ;
\draw (449,205) node [anchor=north west][inner sep=0.75pt]  [rotate=-90] [align=left] {...};
\draw (382,125) node [anchor=north west][inner sep=0.75pt]  [font=\footnotesize]  {$\rho =-i\pi \ $};
\draw (382,185) node [anchor=north west][inner sep=0.75pt]  [font=\footnotesize]  {$\rho =-3i\pi \ $};
\draw (410,50) node [anchor=north west][inner sep=0.75pt]  [font=\footnotesize]  {$\rho =i\pi \ $};
\draw (410,80) node [anchor=north west][inner sep=0.75pt]  [font=\footnotesize]  {$\rho =0\ $};
\draw (345,12) node [anchor=north west][inner sep=0.75pt]  [font=\small]  {$\rho$ - PLANE};
\draw (150,12) node [anchor=north west][inner sep=0.75pt]  [font=\small]  {$\omega$ - PLANE};
\draw (510,103) node [anchor=north west][inner sep=0.75pt]  [font=\footnotesize,color=blue  ,opacity=1 ]  {$\color{blue}\mathcal{C}_{\rm n}$};
\draw (490,63) node [anchor=north west][inner sep=0.75pt]  [font=\footnotesize,color=orange  ,opacity=1 ]  {$\color{orange}\mathcal{C}_{\rm long}$};
\draw (488,185) node [anchor=north west][inner sep=0.75pt]  [font=\footnotesize,color=red  ,opacity=1 ]  {${\color{red} \mathcal{C}_{\rm conv}}$};
\draw (470,144) node [anchor=north west][inner sep=0.75pt]  [font=\small] [align=left] {\begin{minipage}[lt]{21.066468pt}\setlength\topsep{0pt}
\begin{center}{\tiny SADDLE}\end{center}\end{minipage}};
\draw (470,151) node [anchor=north west][inner sep=0.75pt]  [font=\small] [align=left] {\begin{minipage}[lt]{21.066468pt}\setlength\topsep{0pt}
\begin{center}{\tiny POINTS}\end{center}\end{minipage}};
\draw (360,80) node [anchor=north west][inner sep=0.75pt]  [font=\footnotesize,color=green  ,opacity=1 ]  {$\color{darkgreen}\mathcal{C}_{\rm s-left}$};
\draw (210,105) node [anchor=north west][inner sep=0.75pt]  [font=\footnotesize,color=blue  ,opacity=1 ]  {$\color{blue}\mathcal{C}_{\rm n}$};
\draw (235,80) node [anchor=north west][inner sep=0.75pt]  [font=\footnotesize]  {$\omega =1\ $};
\draw (160,125) node [anchor=north west][inner sep=0.75pt]  [font=\footnotesize]  {$\omega =0\ $};
\draw (180,65) node [anchor=north west][inner sep=0.75pt]  [font=\footnotesize,color=red  ,opacity=1 ]  {$\color{red}\mathcal{C}_{\rm conv}$};
\draw (250,220) node [anchor=north east][inner sep=0.75pt]   [align=left] {(a)};
\draw (410,220) node [anchor=north west][inner sep=0.75pt]   [align=left] {(b)}; 
\filldraw (180,100) circle (2pt);
\filldraw [color=gray] (250,100) circle (2pt);
\filldraw (440,70) circle (2pt);
\filldraw [color=gray](440,100) circle (2pt);
\filldraw (440,130) circle (2pt);
\filldraw [color=gray](440,160) circle (2pt);
\filldraw (440,190) circle (2pt);
\end{tikzpicture}
\caption{In (a) we see the $w$-plane. The boundary is at $w=0$ and the horizon at $w=1$. In blue, we see the naive contour $ {\cal C}_{\rm n}$. In red, we have depicted the integration contour ${\cal C}_{\rm conv}$ that leads to a convergent answer when Im$(h)<0$. The singularity is at $w =\infty$. In (b) we see the $\rho$-plane. $\rho=0$ is the horizon and there are multiple images of the singularity at $\rho =  (1+2n)i\pi$. The convergent contour is depicted in red. It can be deformed to the steepest-descent contours shown in green. These pass through the saddle points. We are also left with the contour  ${\cal C}_{\rm s-left}$, from the short geodesic contribution. Additionally, we have ${\cal C}_{\rm long }$ computing the long geodesic contribution, shown in orange. Along the short and long contributions, the propagator takes the form $e^{h \rho}$ and $e^{-h\rho}$, respectively.}
\label{ContPlanar}
\end{center}
\end{figure}

We set 
\be 
 {\rm Im}(h) < 0.
\ee
Then the convergent contour is one that is tilted towards the negative imaginary direction by an angle greater than the angle of the complex number $ -  i  h^*$. In other words, we start with a contour which begins at $\rho=0$ and goes along 
$\rho = -i h^*(1-i \epsilon) \sigma $, with $\sigma \gg 1$ in the complex plane for large $|\rho|$. See figure \ref{ContPlanar}(b). 

Let us return to the full integral. The choice of variables \nref{Choice} is such that the exponent in the propagator is simple
\bea 
\la{Propa}
G &\propto& e^{ - m \ell_{\rm short}(\rho)} + e^{ - m \ell_{\rm long}(\rho) } 
\propto 
e^{ - m \ell_{\rm hor} } \left[ F_{\rm short}(\rho) e^{ h \rho } + F_{\rm long}(\rho) e^{ - h \rho} \right]\,,
\eea 
where
\bea 
\ell_{\rm short}(\rho) = \ell_{hor} - {R \over d } \rho
~,~~~~{\rm and}~~~~
\ell_{\rm long}(\rho) = \ell_{\rm hor} + { R \over d } \rho \,.
\eea 
%are the two terms corresponding to the long and short geodesic contributions. 
The prefactors, $F(\rho)$, in \nref{Propa} do not have exponential dependence on $m$ and are discussed in appendix \ref{Prefactor}.  For now, we will ignore them. 
 
The square of the Weyl tensor \eqref{AdS5Weyl} is 
\be \la{Sour} W^2 \propto 
%=\frac{d(d-1)^2(d-2)}{R^4}
 w^2 \propto { 1 \over ( \cosh{ \rho \over 2 } )^{4} }  = 
 \exp \left[ - 4 \log \left(  \cosh{ \rho \over 2 } \right) \right].
  \ee 
In order to make this term competitive with the propagator term, we can replace $W^2 \to W^{ 2 k}$.
%We could replace $W^2 \to W^{ 2 k}$ to enhance the size of this term and make it competitive to the propagator term. 
For $h<2k$, there is a saddle point along the original integration contour, for real and positive $\rho$. However, we are really interested in the case where $h\gg k$. Notice that \nref{Sour} diverges at 
\be 
\rho = -(1+2n)\,i\pi ,~~~~~~~ n \in \mathbb{Z}.
\ee
In fact, we find  saddle points at 
\be \la{Spva}
\partial_\rho \left( h \rho - m \ell_{\rm hor}   - 4 \log\left( \cosh{\rho\over 2 }\right) \right)=0~~~~ \longrightarrow ~~~~ \rho = - (1+2n)\, \pi i  + \eta  ~,
\ee
where $\eta$ is a small quantity, with a positive real part, given explicitly by
\be
\tanh { \eta \over 2 } = { 2 \over  h }.
\ee

The original tilted contour, ${\cal C}_{\rm conv}$ in figure \ref{ContPlanar}, can be rotated clockwise to the negative real-axis direction. In doing so, this integral can be expressed as a sum of steepest-descent contours passing through the saddle points \nref{Spva}.\footnote{A further derivative of the left most expression in \nref{Spva}  is close to positive at the saddle point \nref{Spva} when $h$ is close to real. This means that   the steepest descent contour indeed goes vertically through the saddle points as in figure \nref{ContPlanar}(b).} All saddle points contribute equally, except for a factor of $e^{-2i\pi h}$. The sum is then proportional to 
\be \la{FSu}
\langle O \rangle \propto 4^{-h} \sum_{n=0}^\infty e^{-(1+ 2 n)\, i\pi h} 
  \propto 4^{-h} { e^{ -i \pi h } \over 1 - e^{ - 2 i\pi h} } \propto { 4^{-h} \over \sin \pi h } ~.
\ee 
In evaluating the exponent, we have only kept the leading term in the large-$h$ expansion and are ignoring powers of $h$ in \nref{FSu}.
If $h$ has a negative imaginary part, higher-order terms in this sum are more and more suppressed. However, it is interesting that they sum up to the inverse sine that we had in the exact answer \nref{IntRe}. 
The overall factor of $4^{-h} = 4^{-\Delta /d }  $ in \nref{FSu} comes from the regularized distance from the boundary to the horizon $\ell_{\rm hor}$ in \nref{RenLe}.

The integral of the short geodesic along the negative real axis, labeled ${\cal C}_{\rm s-left}$ in figure \ref{ContPlanar}, has a similar form to that of the long geodesics and thus, could cancel. To verify this, we need to   compute the prefactors in \nref{Propa} and check that they indeed cancel, see appendix \ref{Prefactor}.   These prefactors have additional singularities and we have not fully understood their effects, see appendix \ref{Prefactor} for a longer discussion.\footnote{The bottom line is that we have only rigorously derived the first saddle $n=0$ in \nref{FSu}, but not the rest, $n> 0$ in \nref{FSu}.}

In summary, we began by considering a black brane. We computed the exact answer by doing the explicit integral in \nref{IntRe}. We considered the geodesic approximation in section \ref{GeoAp}. 
We further justified this approximation through a more detailed saddle-point analysis in section \ref{SaddleAn}, which explained why the contour passes through the saddle point.
We also came upon the added benefit of finding subleading saddles that sum to a $1/\sin (\pi h)$ factor. A similar procedure for the Veneziano amplitude was discussed in appendix A of \cite{Mizera:2019vvs}.
   
\section{More general Schwarzschild black holes} 

Here we will explore the case of more general black holes. A simple generalization is to consider an AdS$_{d+1}$ Schwarzschild black holes with a spherical boundary. These have metrics of the form
\bea  \la{SchAdSd}
ds^2 &=& R^2 \left( -f(r) dt^2 + { dr^2 \over f(r) } + r^2 d\Omega_{d-1}^2 \right)\, ,  \\
f(r) &=& r^2 + 1 -{ \mu \over r^{d-2} } \, .
\eea 
The black brane case is recovered in the $\mu \to \infty $ limit. 

We have not been able to solve the wave equation analytically in this case. In principle, one could do a careful saddle-point analysis. Instead of doing this, we note that in our previous example, a crucial point was to understand the  proper distances to the singularity. We again define $\hat \rho$ in terms of the proper distance\footnote{This is normalized slightly differently than the $\rho$ variables we previously introduced, $\hat \rho = \rho/d$.}
\be \la{PropDi}
d\hat \rho = { dr \over \sqrt{f(r)}}\,,
\ee 
and set $\hat \rho=0$ at the horizon.
For large $\hat \rho$, we have a discussion similar to the one before in the sense that in order to make the integral convergent, we pick out a tilted contour, which spirals to infinity in the $r$-plane. A new feature is that we now have poles at $\Delta = 2 d + 2 n$ from operators like $T \partial ^{2n} T $.  These had vanishing vacuum expectation values for black branes, but not for this more general case \cite{Gobeil:2018fzy}. These operators, in combination with the previous ones, $T^{n}$, give poles at $\Delta=n$ for odd $d$ and $\Delta = 2n$ for even $d$. 

Again, we expect to move the contour into the negative $\hat \rho$ direction and through this process, we expect to pick out saddles near $r=0$. The single point $r=0$ corresponds to many points in $\hat \rho$. We will not figure out the precise structure of the covering space where $\hat \rho$ lives, but we will compute some of the leading values of $\hat \rho$.  
These can be obtained by integrating \nref{PropDi} along various contours. 
One approach for this is to consider a contour in the $r$-plane that starts at the horizon and gets to $r=0$ in various ways.  For example, see figure \ref{rContours}. This is not a real substitute for a full steepest-descent analysis, but it provides us with some information about what to expect. 

\subsection{Four-dimensional black holes}

\begin{figure}[h]
\begin{center}
\vspace{-5mm}
\begin{tikzpicture}[x=0.75pt,y=0.75pt,yscale=-1.25,xscale=1.25]
\draw [color=orange  ,draw opacity=1 ]   (427.75,150) -- (480.2,150) ;
\draw [shift={(453.87,149.76)}, rotate = 539.48] [fill=orange  ,fill opacity=1 ][line width=0.08]  [draw opacity=0] (7.14,-3.43) -- (0,0) -- (7.14,3.43) -- (4.74,0) -- cycle    ;
\draw  [color=orange  ,draw opacity=1 ] (480,150) .. controls (480,149.69) and (480,149.84) .. (480,150) .. controls (480,228.7) and (412.84,292.5) .. (330,292.5) .. controls (247.16,292.5) and (180,228.7) .. (180,150) .. controls (180,149.84) and (180,149.69) .. (180,150) ;
\draw [color=orange  ,draw opacity=1 ]   (179.8,150) -- (260,150) ;
\draw [shift={(220,149.76)}, rotate = 180.34] [fill=orange  ,fill opacity=1 ][line width=0.08]  [draw opacity=0] (7.14,-3.43) -- (0,0) -- (7.14,3.43) -- (4.74,0) -- cycle    ;
\draw [color=orange  ,draw opacity=1 ] [dash pattern={on 4.5pt off 4.5pt}]  (260,150) -- (330,150) ;
\draw [color=darkgreen  ,draw opacity=1 ]   (240,230) .. controls (219.3,210.1) and (239.3,180.1) .. (260,170) ;
\draw [shift={(234.78,195.65)}, rotate = 470.67] [fill=darkgreen  ,fill opacity=1 ][line width=0.08]  [draw opacity=0] (7.14,-3.43) -- (0,0) -- (7.14,3.43) -- (4.74,0) -- cycle    ;
\draw [color=darkgreen  ,draw opacity=1 ]   (239.9,229.9) .. controls (279.3,260.1) and (409.9,180.7) .. (427.75,150) ;
\draw [shift={(341.92,212.8)}, rotate = 333.82] [fill=darkgreen  ,fill opacity=1 ][line width=0.08]  [draw opacity=0] (7.14,-3.43) -- (0,0) -- (7.14,3.43) -- (4.74,0) -- cycle    ;
\draw [color=darkgreen  ,draw opacity=1 ] [dash pattern={on 4.5pt off 4.5pt}]  (260,170) -- (330,150) ;
\draw [shift={(295,160)}, rotate = 524.05] [fill=darkgreen  ,fill opacity=1 ][line width=0.08]  [draw opacity=0] (7.14,-3.43) -- (0,0) -- (7.14,3.43) -- (4.74,0) -- cycle    ;
\draw  [color=tpurple  ,draw opacity=1 ] (427.74,150) .. controls (427.33,222.18) and (378.74,280.6) .. (318.87,280.6) .. controls (258.74,280.6) and (210,221.67) .. (210,148.98) .. controls (210,148.84) and (210,148.7) .. (210,148.56) ;
\draw [color=tpurple  ,draw opacity=1 ]   (330,150) .. controls (329.5,19.9) and (210.5,19.9) .. (210,150) ;
\draw [shift={(317,280.6)}, rotate = 0] [fill=tpurple  ,fill opacity=1 ][line width=0.08]  [draw opacity=0] (7.14,-3.43) -- (0,0) -- (7.14,3.43) -- (4.74,0) -- cycle    ;
\draw [color=blue  ,draw opacity=1 ]   (330,150) .. controls (367.02,170.59) and (388.26,170.59) .. (427.75,150) ;
\draw [shift={(378.84,165.44)}, rotate = 1.79] [fill=blue  ,fill opacity=1 ][line width=0.08]  [draw opacity=0] (5.36,-2.57) -- (0,0) -- (5.36,2.57) -- (3.56,0) -- cycle    ;
\draw   (265,225) -- (255,215)(265,215) -- (255,225);
\draw   (265,85) -- (255,75)(265,75) -- (255,85);
\draw   (432.75,155) -- (422.75,145)(432.75,145) -- (422.75,155);
\draw [line width=0.75]    (330,150) .. controls (331.67,148.33) and (333.33,148.33) .. (335,150) .. controls (336.67,151.67) and (338.33,151.67) .. (340,150) .. controls (341.67,148.33) and (343.33,148.33) .. (345,150) .. controls (346.67,151.67) and (348.33,151.67) .. (350,150) .. controls (351.67,148.33) and (353.33,148.33) .. (355,150) .. controls (356.67,151.67) and (358.33,151.67) .. (360,150) .. controls (361.67,148.33) and (363.33,148.33) .. (365,150) .. controls (366.67,151.67) and (368.33,151.67) .. (370,150) .. controls (371.67,148.33) and (373.33,148.33) .. (375,150) .. controls (376.67,151.67) and (378.33,151.67) .. (380,150) .. controls (381.67,148.33) and (383.33,148.33) .. (385,150) .. controls (386.67,151.67) and (388.33,151.67) .. (390,150) .. controls (391.67,148.33) and (393.33,148.33) .. (395,150) .. controls (396.67,151.67) and (398.33,151.67) .. (400,150) .. controls (401.67,148.33) and (403.33,148.33) .. (405,150) .. controls (406.67,151.67) and (408.33,151.67) .. (410,150) .. controls (411.67,148.33) and (413.33,148.33) .. (415,150) .. controls (416.67,151.67) and (418.33,151.67) .. (420,150) .. controls (421.67,148.33) and (423.33,148.33) .. (425,150) -- (427.75,150) -- (427.75,150) ;
\draw [line width=0.75]    (260,80) .. controls (261.67,81.67) and (261.67,83.33) .. (260,85) .. controls (258.33,86.67) and (258.33,88.33) .. (260,90) .. controls (261.67,91.67) and (261.67,93.33) .. (260,95) .. controls (258.33,96.67) and (258.33,98.33) .. (260,100) .. controls (261.67,101.67) and (261.67,103.33) .. (260,105) .. controls (258.33,106.67) and (258.33,108.33) .. (260,110) .. controls (261.67,111.67) and (261.67,113.33) .. (260,115) .. controls (258.33,116.67) and (258.33,118.33) .. (260,120) .. controls (261.67,121.67) and (261.67,123.33) .. (260,125) .. controls (258.33,126.67) and (258.33,128.33) .. (260,130) .. controls (261.67,131.67) and (261.67,133.33) .. (260,135) .. controls (258.33,136.67) and (258.33,138.33) .. (260,140) .. controls (261.67,141.67) and (261.67,143.33) .. (260,145) .. controls (258.33,146.67) and (258.33,148.33) .. (260,150) .. controls (261.67,151.67) and (261.67,153.33) .. (260,155) .. controls (258.33,156.67) and (258.33,158.33) .. (260,160) .. controls (261.67,161.67) and (261.67,163.33) .. (260,165) .. controls (258.33,166.67) and (258.33,168.33) .. (260,170) .. controls (261.67,171.67) and (261.67,173.33) .. (260,175) .. controls (258.33,176.67) and (258.33,178.33) .. (260,180) .. controls (261.67,181.67) and (261.67,183.33) .. (260,185) .. controls (258.33,186.67) and (258.33,188.33) .. (260,190) .. controls (261.67,191.67) and (261.67,193.33) .. (260,195) .. controls (258.33,196.67) and (258.33,198.33) .. (260,200) .. controls (261.67,201.67) and (261.67,203.33) .. (260,205) .. controls (258.33,206.67) and (258.33,208.33) .. (260,210) .. controls (261.67,211.67) and (261.67,213.33) .. (260,215) .. controls (258.33,216.67) and (258.33,218.33) .. (260,220) -- (260,220) ;
\filldraw[red] (330,150) circle (1.5pt);
\draw (408.75,129.4) node [anchor=north west][inner sep=0.75pt]  [font=\footnotesize]  {$r_{+}$};
\draw (271,202.4) node [anchor=north west][inner sep=0.75pt]  [font=\footnotesize]  {$r_{2} =r^{*}_{1}$};
\draw (271,62.4) node [anchor=north west][inner sep=0.75pt]  [font=\footnotesize]  {$r_{1}$};
\draw (430.75,159) node [anchor=north west][inner sep=0.75pt]  [font=\tiny] [align=left] {HORIZON};
\draw (337,132) node [anchor=north west][inner sep=0.75pt]  [font=\tiny,color={rgb, 255:red, 255; green, 0; blue, 31 }  ,opacity=1 ] [align=left] {\textcolor[rgb]{1,0,0.12}{SINGULARITY}};
\draw (428.75,226.4) node [anchor=north west][inner sep=0.75pt]  [font=\footnotesize,color=orange  ,opacity=1 ]  {$\color{orange} \widetilde{\mathcal{C}}_{1}$};
\draw (358.75,205.4) node [anchor=north west][inner sep=0.75pt]  [font=\footnotesize,color=darkgreen  ,opacity=1 ]  {$\color{darkgreen} \mathcal{C}_{1}$};
\draw (311,54.4) node [anchor=north west][inner sep=0.75pt]  [font=\footnotesize,color=tpurple  ,opacity=1 ]  {$\color{tpurple}\mathcal{C}_{2}$};
\draw (354.75,165.4) node [anchor=north west][inner sep=0.75pt]  [font=\footnotesize,color=blue  ,opacity=1 ]  {$\color{blue}\mathcal{C}_{0}$};
\end{tikzpicture}
\caption{ We see the $r$ complex plane and some branch cuts involved in the definition of $\sqrt{f}$. Integrating \nref{PropDi} along the contours indicated, we get the proper distance to the singularity along various contours. We expect that all of these contribute. Contours ${\cal C}_1$ and $\widetilde{\cal C}_1$ give the same answer, but the second one is convenient to obtain \nref{SecRoot}.  }
\label{rContours}
\end{center}
\end{figure}
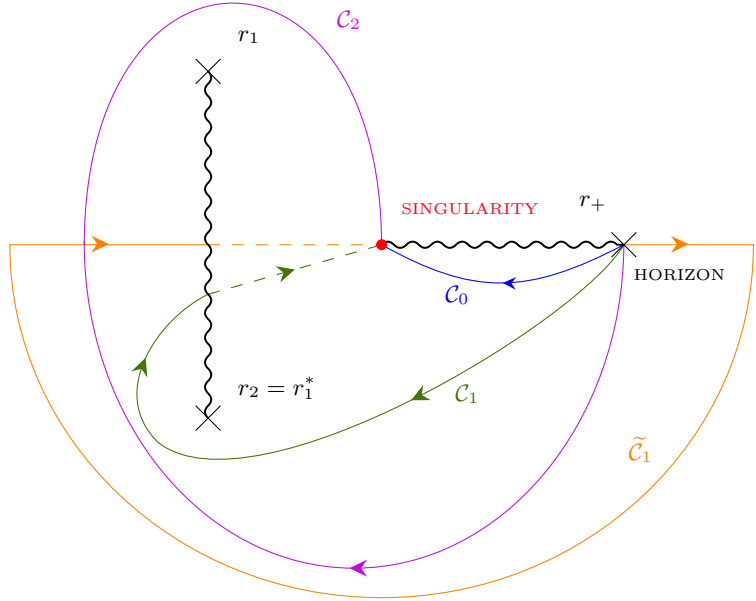

As a first example, let us consider the case of $d=3$, corresponding to a black hole in AdS$_4$. 
We write $f(r)$ in \nref{SchAdSd} as 
\be 
f(r) = { (r-r_+) (r -r_1) (r-r_1^*) \over r}
~,~~~~~ r_1 = - {r_+ \over 2} + i \sqrt{ 1 + {3 \over 4}r_+^2}
~,~~~~~
\mu = r_+ (1+r_+^2)\,.
\ee
When we consider $1/\sqrt{f(r)}$, we can run the branch cuts as indicated in figure \ref{rContours}. The convergent contour for the short geodesic   spirals clockwise around the complex $r$-plane for large values of $r$. As we try to deform the contour into the region of a decreasing propagator, we want to go under all these branch cuts. We will not analyze exactly how to do so. However, we notice that as we move the contour, we will encounter the singularities in $W^2$ at $r=0$. It is interesting to find out where they first occur in the $\hat \rho$-plane. Setting the origin of the $\hat \rho$-plane at the horizon, we find that the first singularity occurs at 
 \be 
 \hat \rho_0 =- i \chi_0  ~,~~~~~~ \chi_0 \equiv \int_0^{r_+} { dr \over \sqrt{-f(r)}}\,,
 \ee
and corresponds to the integral along the contour ${\cal C}_0$ in figure \ref{rContours}. 
We can also reach $r=0$ following the contour ${\cal C}_1$ in figure \ref{rContours}, giving 
\be 
 \hat \rho_1 = - i\pi + \gamma 
~,~~~~~ \gamma = \lim_{r_{\rm c} \to \infty }
 \left[  \int_{r_+}^{r_{\rm c}}  { dr \over \sqrt{f(r)}} - \int_{-r_{\rm c}}^0 { d r \over \sqrt{f(r)} }  \right] \, .
\la{SecRoot}
\ee
  This formula is derived by deforming the integral to $\widetilde {\cal C}_1$. We note that this $\gamma$ is positive. 
%We could also consider the contour ${\cal C}_1'$ in figure \ref{rContours}, which gives 
%\be 
%\rho_1' = - \pi i - \gamma ~.
%\ee
We can also consider the contour ${\cal C}_2$ in figure \ref{rContours}:
\begin{equation}
\hat \rho_2 =  i \left(\chi_0 - 2\pi\right)\, .
\end{equation}
Additionally, we can add $- 2i\pi $ to all of the above by circling more times around infinity. We are not sure if these are all of them, or if all of these do indeed contribute. 

However, if the imaginary part of $\Delta $ is large, then only $\hat \rho_0$ dominates. On the other hand, if the imaginary part of $\Delta $ is very small, then the one involving $\hat \rho_1$ dominates, since it has the largest real part. 
Thus, the question of whether or not the information about the proper time to the singularity is contained in the leading term depends on the size of ${\rm Im}(h)$. 

The schematic form of the answer is 
\be  \la{AnsMas}
\langle O \rangle  \sim  e^{ - m \ell_{\rm hor} } \left( { a e^{- i   \chi_0\Delta   } +  b e^{-i( \pi +i \gamma) \Delta  } 
%+b' e^{ ( -i \pi - \gamma) \Delta } 
+ a^* e^{- i( 2 \pi -\chi_0)\Delta } + \cdots  \over 1 - e^{ -2 i\pi \Delta }   }\right) \,,
\ee
where the dots indicate possible further 
exponentials. The denominator comes from the circles around $r=0$.  The prefactors $a$, $b$ have only power-law dependence on $h$. The overall factor in \nref{AnsMas} comes from the renormalized distance from infinity to the horizon and is equal to 
\be \la{ellpl}
\ell_{\rm hor} =    R \lim_{r_{\rm c} \to \infty} \left[ \int_{r_+}^{r_{\rm c}} { dr \over \sqrt{f(r)} }  - \log r_{\rm c} \right].
\ee
 
As a check, when we go to the black brane limit, $r_+ \to \infty$, we find that $\chi_0 \sim \pi/3$ and $\gamma \sim  0$~\footnote{Numerically,  we found that  we approach these values as $\chi = \pi/3 - 0.25/r_+^2 $ and $\gamma = 0.43/r_+^2$ when $r_+ \to \infty$.}
, so that the exponents in \nref{AnsMas} become 
$e^{ -i\pi \Delta/3}$, $e^{ -i\pi\Delta}$, $e^{ -5 i\pi  \Delta/3}$. We further expect that $a$ becomes equal to $b$, so that  
\be 
\langle O \rangle_{\mu \to \infty} \propto  e^{-m \ell_{\rm hor}} \left( 
{ e^{-i\pi \Delta/3} ( 1 + q + q^2) \over (1 -q^3) } \right) = z_0^{-\Delta} 4^{-\Delta/3}  { e^{ -i\pi \Delta/3} \over ( 1 - e^{ -2 i\pi\Delta/3} ) } %~,~~~~~q = e^{- 2 \pi i {\Delta / 3} },
\, ,
\ee 
where above we have used the large-$r_+$ value of $\ell_{\rm hor}$ \nref{ellpl} and defined $q\equiv e^{- 2 i\pi {\Delta / 3} }$. We see that we reproduce the black brane result \nref{FSu} for $d=3$, with $z_0 \sim 1/r_+$. 

Let us make some comments: 
\begin{itemize}
	\item The poles in \nref{AnsMas} are, in general, at integer values $\Delta =n$. This comes from the combination of operators $T^{3n}$ (for $d=3$) and $T  \partial ^{ 2 m } T $. In the black-brane limit, the latter have zero expectation value and only the poles at $\Delta = 3 n$ are present. 
	\item 
	The positions of the poles are determined by the operator content of the theory. In fact, the basic spacing is set by the integral of a full circle at large values of $r$, which is fixed by the form of the theory near the boundary. 
	\item 
	However, the numerator (i.e. the position of the zeros) depends on the details of the black hole and its temperature. For large $r_+$, we find that they precisely cancel some of the poles. 
	\item
	In the small mass limit $\left(r_+ \to 0\right)$, we get $\chi_0 \sim \pi r_+/2$, which sets the time to the singularity for a black hole in flat space. In this regime we get $\gamma \sim r_+ (-\log r_+) $. This logarithmic divergence is interpreted as coming from the region outside the black hole, where AdS is approximated by flat space.  We expect the effects of the second root $\hat \rho_1$ \nref{SecRoot} to be reflective of the contribution \textit{not} from the interior of the flat space black hole, but rather from the region near the center of AdS and outside the black hole. We have not yet understood this in detail.
\end{itemize}

\subsection{Five-dimensional black holes}

Here we insert $d=4$ in \nref{SchAdSd}. After redefining $u\equiv r^2$, we find that the proper length involves
\be 
d\hat \rho = { 1 \over 2}   { du \over \sqrt{ (u-u_+) (u + u_+ + 1) } }
~~~~\longrightarrow ~~~~
{ u - u_+ \over 1 + 2 u_+} = \sinh^2 \hat \rho\, ,
\ee 
where $u_+$ parametrizes the position of the horizon. 
\begin{figure}[h]
\begin{center}
\hspace{-10mm}
\begin{tikzpicture}[x=0.75pt,y=0.75pt,yscale=-1.5,xscale=1.5] 
\draw [color=tpurple  ,draw opacity=1 ]   (200,65) .. controls (200.52,25.7) and (340.52,26.8) .. (340,65);
\draw [color=tpurple  ,draw opacity=1 ]   (200,65) .. controls (200.52,105.7) and (400.52,124.8) .. (440,65);
\draw [shift={(320.32,103.02)}, rotate = 0.61] [fill=tpurple  ,fill opacity=1 ][line width=0.08]  [draw opacity=0] (5.36,-2.57) -- (0,0) -- (5.36,2.57) -- (3.56,0) -- cycle;
\draw [shift={(270.39,35.94)}, rotate = 539.27] [fill=tpurple  ,fill opacity=1 ][line width=0.08]  [draw opacity=0] (5.36,-2.57) -- (0,0) -- (5.36,2.57) -- (3.56,0) -- cycle;
\draw [color=blue  ,draw opacity=1 ]   (340,65) .. controls (377.02,85.59) and (400.52,85.59) .. (440,65);
\draw [shift={(390,80.44)}, rotate = 1.72] [fill=blue  ,fill opacity=1 ][line width=0.08]  [draw opacity=0] (5.36,-2.57) -- (0,0) -- (5.36,2.57) -- (3.56,0) -- cycle;
\draw    (440,65) .. controls (438.33,66.67) and (436.67,66.67) .. (435,65) .. controls (433.33,63.33) and (431.67,63.33) .. (430,65) .. controls (428.33,66.67) and (426.67,66.67) .. (425,65) .. controls (423.33,63.33) and (421.67,63.33) .. (420,65) .. controls (418.33,66.67) and (416.67,66.67) .. (415,65) .. controls (413.33,63.33) and (411.67,63.33) .. (410,65) .. controls (408.33,66.67) and (406.67,66.67) .. (405,65) .. controls (403.33,63.33) and (401.67,63.33) .. (400,65) .. controls (398.33,66.67) and (396.67,66.67) .. (395,65) .. controls (393.33,63.33) and (391.67,63.33) .. (390,65) .. controls (388.33,66.67) and (386.67,66.67) .. (385,65) .. controls (383.33,63.33) and (381.67,63.33) .. (380,65) .. controls (378.33,66.67) and (376.67,66.67) .. (375,65) .. controls (373.33,63.33) and (371.67,63.33) .. (370,65) .. controls (368.33,66.67) and (366.67,66.67) .. (365,65) .. controls (363.33,63.33) and (361.67,63.33) .. (360,65) .. controls (358.33,66.67) and (356.67,66.67) .. (355,65) .. controls (353.33,63.33) and (351.67,63.33) .. (350,65) .. controls (348.33,66.67) and (346.67,66.67) .. (345,65) .. controls (343.33,63.33) and (341.67,63.33) .. (340,65) -- (340,65) ;
\draw    (340,65) .. controls (338.33,66.67) and (336.67,66.67) .. (335,65) .. controls (333.33,63.33) and (331.67,63.33) .. (330,65) .. controls (328.33,66.67) and (326.67,66.67) .. (325,65) .. controls (323.33,63.33) and (321.67,63.33) .. (320,65) .. controls (318.33,66.67) and (316.67,66.67) .. (315,65) .. controls (313.33,63.33) and (311.67,63.33) .. (310,65) .. controls (308.33,66.67) and (306.67,66.67) .. (305,65) .. controls (303.33,63.33) and (301.67,63.33) .. (300,65) .. controls (298.33,66.67) and (296.67,66.67) .. (295,65) .. controls (293.33,63.33) and (291.67,63.33) .. (290,65) .. controls (288.33,66.67) and (286.67,66.67) .. (285,65) .. controls (283.33,63.33) and (281.67,63.33) .. (280,65) .. controls (278.33,66.67) and (276.67,66.67) .. (275,65) .. controls (273.33,63.33) and (271.67,63.33) .. (270,65) .. controls (268.33,66.67) and (266.67,66.67) .. (265,65) .. controls (263.33,63.33) and (261.67,63.33) .. (260,65) .. controls (258.33,66.67) and (256.67,66.67) .. (255,65) .. controls (253.33,63.33) and (251.67,63.33) .. (250,65) .. controls (248.33,66.67) and (246.67,66.67) .. (245,65) .. controls (243.33,63.33) and (241.67,63.33) .. (240,65) -- (240,65) ;
\draw   (434.8,59.7) -- (445.41,70.3)(445.41,59.7) -- (434.8,70.3) ;
\draw   (234.8,59.7) -- (245.41,70.3)(245.41,59.7) -- (234.8,70.3) ;
\filldraw[red] (340,65) circle (1.5pt);
\draw (329,74.4) node [anchor=north west][inner sep=0.75pt]  [font=\scriptsize]  {$u=0$};
\draw (374,82.4) node [anchor=north west][inner sep=0.75pt]  [font=\scriptsize,color=blue  ,opacity=1 ]  {$\color{blue}\mathcal{C}_{0}$};
\draw (442,72.4) node [anchor=north west][inner sep=0.75pt]  [font=\scriptsize]  {$u_{+}$};
\draw (211,72.4) node [anchor=north west][inner sep=0.75pt]  [font=\scriptsize]  {$-1-u_{+}$};
\draw (341,47) node [anchor=north west][inner sep=0.75pt]  [font=\tiny,color={rgb, 255:red, 255; green, 0; blue, 31 }  ,opacity=1 ] [align=left] {\textcolor{red}{SINGULARITY}};
\draw (244,22.4) node [anchor=north west][inner sep=0.75pt]  [font=\scriptsize,color=tpurple  ,opacity=1 ]  {$\color{tpurple}\mathcal{C}_{1}$};
\end{tikzpicture}
\vspace{-5mm}
\caption{We see the integration contours in the $u$-plane that define $\hat \rho_0$ and $\hat \rho_1$.}
\label{FivedContours}
\end{center}
\end{figure}
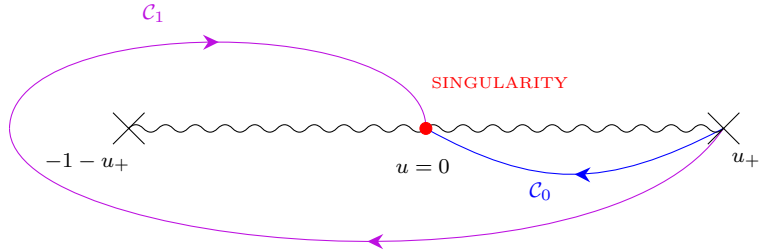

In this case, it is very easy to find the $\hat \rho$ positions of the $r=u=0$ regions. They sit at
\be 
\hat \rho_0 = -i \chi_0  ~,~~~~~~\hat \rho_1 = - i (\pi - \chi_0)  ~,~~~~~~~ \sin^2 \chi_0 = { u_+ \over 1 + 2 u_+} \,,
\ee 
and we can add any multiple of $-i\pi$ to these values.  
Here we see that for large $u_+$, we get $\chi_0 = \pi/4$. In this case then, we expect that these saddles combine as 
\be \la{oneptFive}
\langle O \rangle
~~~\sim ~~~
e^{ - m \ell_{\rm hor} }  e^{- i \pi \Delta/4} 
\left({ 1 + e^{ -i \pi \Delta/2} \over 1 - e^{ -i \pi \Delta } } \right)
~~~\sim ~~~
{ 4^{ -\Delta/4}  \over \sin \left({\pi \Delta \over 4}\right)}\,,
\ee
as we had in  \nref{FSu}. Note that 
\be 
{ \ell_{\rm hor} \over R} = \half \lim_{u_{\rm c} \to \infty} \left[ \int_{u_+}^{u_{\rm c}} { d u 
\over \sqrt{ (u-u_+) (u + u_+ + 1) } } - \log u_{\rm c} \right] = - \half \log( 1 + 2 u_+) + \log (2) .
\ee 
Assuming that the prefactors associated with all saddles are real, we find
\be 
\langle O \rangle 
~~\propto ~~
( 1 + 2 u_+)^{ \Delta/2} \, 2^{ -\Delta} \, { \cos ({ \pi \over 2} - \chi_0 ) \over \sin\left({ \pi \Delta \over 2 } \right)}\, .
\ee 

Let us make some comments:
\begin{itemize}
	\item The poles at $\Delta = 2 n $ are what we expect from the operators, $T^{ m}$ and $T  \partial ^{ 2 m} T $, when the dimension of $T$ is even (four in this case). This is slightly different than what we had found for $d=3$, where the poles were at $\Delta =n$.  
	\item 
	In the small-$r_+$  limit, we see that $\chi_0 \to r_+$, which is what we expect for the time to the singularity for the flat-space black hole in five dimensions. 
	\item
	We can also give a physical interpretation to the metric in the $u< 0$ region. This corresponds to replacing the $S^3$ with $H^3$, and considering a hyperbolic black hole\footnote{Recall that under $\theta \to i \rho$, we have that $ds^2 = d\theta^2 + \sin^2 \theta \, d\Omega_2^2 ~~\to~~ ds^2 = - [ d\rho^2 + \sinh^2\rho \, d\Omega_2^2]$. For $u<0$, the $ u \, d\Omega_3^2$ term in the metric is interpreted as  $(-u)\,ds^2_{H_3}$, which now has a positive coefficient.}. Then, the point at $u = - 1 - u_+$ is the horizon and the time to the singularity for this new black hole is $R ({\pi\over 2} - \chi_0)$. 
\end{itemize}

\section{Black holes with an inner horizon} 
\la{InnerSec}

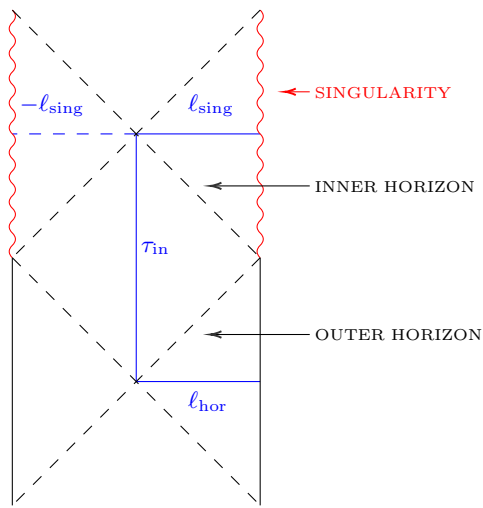
\begin{figure}[h]
\begin{center}
\makebox{
\hspace{23mm}
\begin{tikzpicture}[x=0.75pt,y=0.75pt,yscale=-1.25,xscale=1.25]
\draw [color=blue,draw opacity=1 ]   (330,60) -- (330,160) ;
\draw [color=blue,draw opacity=1 ]   (380,160) -- (330,160) ;
\draw [color=blue,draw opacity=1 ]   (380,60) -- (330,60) ; 
\draw [color=blue,draw opacity=1 ] [dash pattern={on 4.5pt off 4.5pt}]  (330,60) -- (280,60) ; 
\draw [color=red  ,draw opacity=1 ]   (280,10) .. controls (281.67,11.67) and (281.67,13.33) .. (280,15) .. controls (278.33,16.67) and (278.33,18.33) .. (280,20) .. controls (281.67,21.67) and (281.67,23.33) .. (280,25) .. controls (278.33,26.67) and (278.33,28.33) .. (280,30) .. controls (281.67,31.67) and (281.67,33.33) .. (280,35) .. controls (278.33,36.67) and (278.33,38.33) .. (280,40) .. controls (281.67,41.67) and (281.67,43.33) .. (280,45) .. controls (278.33,46.67) and (278.33,48.33) .. (280,50) .. controls (281.67,51.67) and (281.67,53.33) .. (280,55) .. controls (278.33,56.67) and (278.33,58.33) .. (280,60) .. controls (281.67,61.67) and (281.67,63.33) .. (280,65) .. controls (278.33,66.67) and (278.33,68.33) .. (280,70) .. controls (281.67,71.67) and (281.67,73.33) .. (280,75) .. controls (278.33,76.67) and (278.33,78.33) .. (280,80) .. controls (281.67,81.67) and (281.67,83.33) .. (280,85) .. controls (278.33,86.67) and (278.33,88.33) .. (280,90) .. controls (281.67,91.67) and (281.67,93.33) .. (280,95) .. controls (278.33,96.67) and (278.33,98.33) .. (280,100) .. controls (281.67,101.67) and (281.67,103.33) .. (280,105) .. controls (278.33,106.67) and (278.33,108.33) .. (280,110) -- (280,110) ;
\draw [color=red  ,draw opacity=1 ]   (380,10) .. controls (381.67,11.67) and (381.67,13.33) .. (380,15) .. controls (378.33,16.67) and (378.33,18.33) .. (380,20) .. controls (381.67,21.67) and (381.67,23.33) .. (380,25) .. controls (378.33,26.67) and (378.33,28.33) .. (380,30) .. controls (381.67,31.67) and (381.67,33.33) .. (380,35) .. controls (378.33,36.67) and (378.33,38.33) .. (380,40) .. controls (381.67,41.67) and (381.67,43.33) .. (380,45) .. controls (378.33,46.67) and (378.33,48.33) .. (380,50) .. controls (381.67,51.67) and (381.67,53.33) .. (380,55) .. controls (378.33,56.67) and (378.33,58.33) .. (380,60) .. controls (381.67,61.67) and (381.67,63.33) .. (380,65) .. controls (378.33,66.67) and (378.33,68.33) .. (380,70) .. controls (381.67,71.67) and (381.67,73.33) .. (380,75) .. controls (378.33,76.67) and (378.33,78.33) .. (380,80) .. controls (381.67,81.67) and (381.67,83.33) .. (380,85) .. controls (378.33,86.67) and (378.33,88.33) .. (380,90) .. controls (381.67,91.67) and (381.67,93.33) .. (380,95) .. controls (378.33,96.67) and (378.33,98.33) .. (380,100) .. controls (381.67,101.67) and (381.67,103.33) .. (380,105) .. controls (378.33,106.67) and (378.33,108.33) .. (380,110) -- (380,110) ;
\draw  [dash pattern={on 4.5pt off 4.5pt}]  (380,10) -- (280,110) ;
\draw  [dash pattern={on 4.5pt off 4.5pt}]  (280,10) -- (380,110) ;
\draw  [dash pattern={on 4.5pt off 4.5pt}]  (380,110) -- (280,210) ;
\draw  [dash pattern={on 4.5pt off 4.5pt}]  (280,110) -- (380,210) ;
\draw  (280, 110) -- (280,210); 
\draw  (380, 110) -- (380,210);  
\draw  [color = red]  (400,43) -- (390,43) ;
\draw [shift={(390,43)}, rotate = 0] [color=red  ][line width=0.75]    (4.37,-1.32) .. controls (2.78,-0.56) and (1.32,-0.12) .. (0,0) .. controls (1.32,0.12) and (2.78,0.56) .. (4.37,1.32)   ;
\draw    (400,81) -- (360,81) ;
\draw [shift={(360,81)}, rotate = 0] [color=black  ][line width=0.75]    (4.37,-1.32) .. controls (2.78,-0.56) and (1.32,-0.12) .. (0,0) .. controls (1.32,0.12) and (2.78,0.56) .. (4.37,1.32)   ;
\draw    (400,141) -- (360,141) ;
\draw [shift={(360,141)}, rotate = 0] [color=black  ][line width=0.75]    (4.37,-1.32) .. controls (2.78,-0.56) and (1.32,-0.12) .. (0,0) .. controls (1.32,0.12) and (2.78,0.56) .. (4.37,1.32)   ;
\draw (401,40) node [anchor=north west][inner sep=0.75pt]  [font=\tiny,color=red ,opacity=1 ] [align=left] {\textcolor{red}{SINGULARITY}\\};
\draw (331,102.4) node [anchor=north west][inner sep=0.75pt]  [font=\footnotesize,color=blue  ,opacity=1 ]  {$\color{blue}\tau _{\rm in}$};
\draw (350,163) node [anchor=north west][inner sep=0.75pt]  [font=\footnotesize,color=blue  ,opacity=1 ]  {$\color{blue}\ell _{\rm hor}$};
\draw (350,43) node [anchor=north west][inner sep=0.75pt]  [font=\footnotesize,color=blue  ,opacity=1 ]  {$\color{blue}\ell _{\rm sing}$};
\draw (310,43) node [anchor=north east][inner sep=0.75pt]  [font=\footnotesize,color=blue  ,opacity=1 ]  {$\color{blue}-\ell _{\rm sing}$};
\draw (401,78) node [anchor=north west][inner sep=0.75pt]  [font=\tiny] [align=left] {INNER HORIZON};
\draw (401,138) node [anchor=north west][inner sep=0.75pt]  [font=\tiny] [align=left] {OUTER HORIZON};
\end{tikzpicture}}
\caption{Penrose diagram of a charged black hole in AdS. We have both an outer horizon and an inner horizon. The one-point function would involve the length of geodesics roughly as indicated, as well as an imaginary contribution that has the size of the order of the time between the inner and outer horizons. The distance from the inner bifurcation surface to the singularity, $\ell_{\rm sing}$,  also appears.   }
\label{Penrose}
\end{center}
\end{figure}

Here we discuss some aspects of black holes with an inner horizon. We will consider spherical charged black holes. Their Penrose diagram is shown in figure \ref{Penrose}. We are ignoring backreaction, so we do not expect any singularity in the inner horizon. We will present some evidence that, in the large-mass expansion with Im$(m)<0$, the ``phase'' of the one-point function tells us about the time between the outer and inner horizons. The geodesics can be interpreted as going to the left or right in the Penrose diagram, so that schematically they look like they are going to the timelike singularities in figure \ref{Penrose}. 

\begin{figure}[h]
\begin{center}
\begin{tikzpicture}[x=0.75pt,y=0.75pt,yscale=-1.5,xscale=1.5]
\draw [line width=1]    (340,69.06) .. controls (338.33,70.73) and (336.67,70.73) .. (335,69.06) .. controls (333.33,67.39) and (331.67,67.39) .. (330,69.06) .. controls (328.33,70.73) and (326.67,70.73) .. (325,69.06) .. controls (323.33,67.39) and (321.67,67.39) .. (320,69.06) .. controls (318.33,70.73) and (316.67,70.73) .. (315,69.06) .. controls (313.33,67.39) and (311.67,67.39) .. (310,69.06) .. controls (308.33,70.73) and (306.67,70.73) .. (305,69.06) .. controls (303.33,67.39) and (301.67,67.39) .. (300,69.06) .. controls (298.33,70.73) and (296.67,70.73) .. (295,69.06) .. controls (293.33,67.39) and (291.67,67.39) .. (290,69.06) .. controls (288.33,70.73) and (286.67,70.73) .. (285,69.06) .. controls (283.33,67.39) and (281.67,67.39) .. (280,69.06) .. controls (278.33,70.73) and (276.67,70.73) .. (275,69.06) .. controls (273.33,67.39) and (271.67,67.39) .. (270,69.06) .. controls (268.33,70.73) and (266.67,70.73) .. (265,69.06) .. controls (263.33,67.39) and (261.67,67.39) .. (260,69.06) .. controls (258.33,70.73) and (256.67,70.73) .. (255,69.06) .. controls (253.33,67.39) and (251.67,67.39) .. (250,69.06) .. controls (248.33,70.73) and (246.67,70.73) .. (245,69.06) .. controls (243.33,67.39) and (241.67,67.39) .. (240,69.06) -- (240,69.06) ;
\draw [color=orange  ,draw opacity=1 ]   (180,70) .. controls (179.97,130.35) and (419.97,150.29) .. (460,70) ;
\draw [shift={(318.68,123)}, rotate = 0.47] [fill=orange  ,fill opacity=1 ][line width=0.08]  [draw opacity=0] (5.36,-2.57) -- (0,0) -- (5.36,2.57) -- (3.56,0) -- cycle    ;
\draw [color=orange  ,draw opacity=1 ]   (180,70) .. controls (180.06,10.89) and (410.06,10.89) .. (410,70) ;
\draw [shift={(295.05,25.67)}, rotate = 539.56] [fill=orange  ,fill opacity=1 ][line width=0.08]  [draw opacity=0] (5.36,-2.57) -- (0,0) -- (5.36,2.57) -- (3.56,0) -- cycle    ;
\draw [color=orange  ,draw opacity=1 ] [dash pattern={on 4.5pt off 4.5pt}]  (340,70) .. controls (340.31,99.39) and (410.31,99.39) .. (410,70) ;
\draw [shift={(375.76,92.14)}, rotate = 2.04] [fill=orange  ,fill opacity=1 ][line width=0.08]  [draw opacity=0] (5.36,-2.57) -- (0,0) -- (5.36,2.57) -- (3.56,0) -- cycle    ;
\draw [color=darkgreen  ,draw opacity=1 ]   (200,70) .. controls (200.52,109.77) and (421.11,138.19) .. (460,70) ;
\draw [shift={(330.09,110.98)}, rotate = 1.76] [fill=darkgreen  ,fill opacity=1 ][line width=0.08]  [draw opacity=0] (5.36,-2.57) -- (0,0) -- (5.36,2.57) -- (3.56,0) -- cycle    ;
\draw [color=darkgreen  ,draw opacity=1 ]   (200,70) .. controls (200.52,29.77) and (340.52,30.86) .. (340,70) ;
\draw [shift={(270.39,40)}, rotate = 539.27] [fill=darkgreen  ,fill opacity=1 ][line width=0.08]  [draw opacity=0] (5.36,-2.57) -- (0,0) -- (5.36,2.57) -- (3.56,0) -- cycle    ;
\draw [color=blue  ,draw opacity=1 ]   (340,70) .. controls (335.51,100.59) and (433.91,108.99) .. (460,70) ;
\draw [shift={(397.65,96.03)}, rotate = 359.64] [fill=blue  ,fill opacity=1 ][line width=0.08]  [draw opacity=0] (5.36,-2.57) -- (0,0) -- (5.36,2.57) -- (3.56,0) -- cycle    ;
\draw [color=blue  ,draw opacity=1 ]   (340,70) .. controls (349.51,77.39) and (392.31,78.59) .. (400,70) ;
\draw [shift={(370.29,76.09)}, rotate = 1.63] [fill=blue  ,fill opacity=1 ][line width=0.08]  [draw opacity=0] (5.36,-2.57) -- (0,0) -- (5.36,2.57) -- (3.56,0) -- cycle    ;
\draw [color=blue  ,draw opacity=1 ]   (400,70) .. controls (409.51,76.57) and (452.31,78.59) .. (460,70) ;
\draw [shift={(430.29,75.68)}, rotate = 2.53] [fill=blue  ,fill opacity=1 ][line width=0.08]  [draw opacity=0] (5.36,-2.57) -- (0,0) -- (5.36,2.57) -- (3.56,0) -- cycle    ;
\draw [color=tpurple  ,draw opacity=1 ]   (420,70) .. controls (419.51,82.99) and (450.31,84.59) .. (460,70) ;
\draw [shift={(439.14,80.28)}, rotate = 1.51] [fill=tpurple  ,fill opacity=1 ][line width=0.08]  [draw opacity=0] (5.36,-2.57) -- (0,0) -- (5.36,2.57) -- (3.56,0) -- cycle    ; 
\draw [color=tpurple  ,draw opacity=1 ] [dash pattern={on 4.5pt off 4.5pt}]  (340,70) .. controls (340.31,39.39) and (420.31,40.19) .. (420,70) ;
\draw [shift={(380.83,47.23)}, rotate = 358.91] [fill=tpurple  ,fill opacity=1 ][line width=0.08]  [draw opacity=0] (5.36,-2.57) -- (0,0) -- (5.36,2.57) -- (3.56,0) -- cycle    ;
\draw [line width=1]    (390,70) .. controls (391.67,68.33) and (393.33,68.33) .. (395,70) .. controls (396.67,71.67) and (398.33,71.67) .. (400,70) .. controls (401.67,68.33) and (403.33,68.33) .. (405,70) .. controls (406.67,71.67) and (408.33,71.67) .. (410,70) .. controls (411.67,68.33) and (413.33,68.33) .. (415,70) .. controls (416.67,71.67) and (418.33,71.67) .. (420,70) .. controls (421.67,68.33) and (423.33,68.33) .. (425,70) .. controls (426.67,71.67) and (428.33,71.67) .. (430,70) .. controls (431.67,68.33) and (433.33,68.33) .. (435,70) .. controls (436.67,71.67) and (438.33,71.67) .. (440,70) .. controls (441.67,68.33) and (443.33,68.33) .. (445,70) .. controls (446.67,71.67) and (448.33,71.67) .. (450,70) .. controls (451.67,68.33) and (453.33,68.33) .. (455,70) .. controls (456.67,71.67) and (458.33,71.67) .. (460,70) -- (460,70) ;
\draw [line width=1.1]  (235,64) -- (245,74)(245,64) -- (235,74) ;
\draw [line width=1.1]  (385,65) -- (395,75)(395,65) -- (385,75) ;
\draw [line width=1.1]  (455,65) -- (465,75)(465,65) -- (455,75) ;
\filldraw[red] (340,70) circle (1.5pt);
\draw (355,95) node [anchor=north west][inner sep=0.75pt]  [font=\scriptsize,color=blue  ,opacity=1 ]  {$\color{blue}\mathcal{C}_{0}$};
\draw (461,77) node [anchor=north west][inner sep=0.75pt]  [font=\scriptsize]  {$u_{+}$};
\draw (215,76.46) node [anchor=north west][inner sep=0.75pt]  [font=\scriptsize]  {$-1-u_{+} -u_{-}$};
\draw (252,104) node [anchor=north west][inner sep=0.75pt]  [font=\scriptsize,color=darkgreen  ,opacity=1 ]  {$\color{darkgreen}\mathcal{C}_{1}$};
\draw (390,55) node [anchor=north west][inner sep=0.75pt]  [font=\scriptsize]  {$u_{-}$};
\draw (200,108) node [anchor=north west][inner sep=0.75pt]  [font=\scriptsize,color=orange  ,opacity=1 ]  {$\color{orange}\mathcal{C}_{2}$};
\draw (350,40) node [anchor=north west][inner sep=0.75pt]  [font=\scriptsize,color=tpurple  ,opacity=1 ]  {$\color{tpurple}\mathcal{C}_{3}$};
\draw (377,78) node [anchor=north west][inner sep=0.75pt]  [font=\tiny,color=blue  ,opacity=1 ]  {$\color{blue}\gamma_{0}$};
\draw (403,76) node [anchor=north west][inner sep=0.75pt]  [font=\tiny,color=blue  ,opacity=1 ]  {$\color{blue}-i\chi _{0}$};
\end{tikzpicture}
\vspace{-5mm}
\caption{ We depict the $u =r^2$ plane and the various branch points in the computation of $\rho$. The   depicted contours give us various values for the ``distance'' to the singularity, which are expected to contribute to the one-point function. The dotted lines indicate that we go to the second sheet by crossing the cut. We have given two equivalent forms for the contour ${\cal C}_0$ --- one of them emphasizes the origin of the $-i\chi_0$ and $\gamma_0$ contributions. 
%The contour ${\cal C}_0$ is in blue where we also divided it into the two pieces giving $\chi_0$ and $\gamma_0$. ${\cal C}_1$ is in purple. ${\cal C}_2$ in orange. ${\cal C}_3$ in green. 
}
\label{NearExt}
\end{center}
\end{figure}
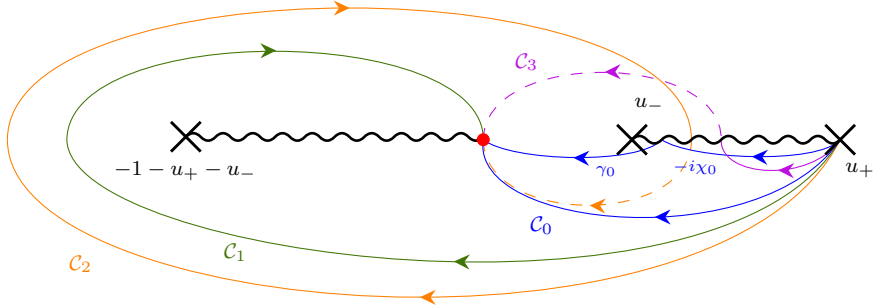

As a simple case, let us consider a five-dimensional charged AdS black hole with the same metric as in \nref{SchAdSd} but with 
\be 
f(r) = r^2 + 1  - { \mu \over r^2 } +  { q^2 \over r^4 } = 
{ (u - u_+) (u-u_-) ( u + 1 + u_+ + u_-) \over u^2 } ~,~~~~~u \equiv r^2\,, 
\ee
where $u_+ > u_- > 0$. Here we have parametrized $\mu$ and $q^2$ in terms of $u_+$ and $u_-$ in the regime where we have a smooth horizon. The structure of branch cuts when we write 
$d\hat \rho = {dr\over\sqrt{f(r)}} = \half {du \over \sqrt{uf(u)}}$ is depicted in figure \ref{NearExt}. 

The structure here is somewhat similar to that of the four-dimensional black hole. However, here all the branch cuts are on the real axis. It is interesting to consider the total length $\hat \rho$ of the $r=u=0$ singularity when we follow the contours indicated in figure \ref{NearExt}. The first value, $\hat \rho_0$, has both real and an imaginary parts:
\bea 
 \hat \rho_0 &=& -i \chi_0 + \gamma_0  ~,~~~~~~ \la{FSa}
\\
\chi_0 &=& \half \int_{w_-}^{w_+} dw {
 \sqrt{ w}  \over \sqrt{(w_+-w)(w-w_-) (w + 1 + w_+ + w_-) } }\,,
 \cr 
 \gamma_0 &=& \half \int_{0}^{w_-} dw{
 \sqrt{ w}  \over \sqrt{(w_+-w)(w_- -w) (w + 1 + w_+ + w_-) } }\,,
\eea
where we have indicated more explicitly the integrals we are considering. 
The imaginary part is given by the time $\chi_0$ between the inner and outer horizons, while the real part contains the distance $\ell_{\rm sing} = R \gamma_0$ between the bifurcation surface of the inner horizon and the singularity (see figure \ref{Penrose}). 
 The saddle \nref{FSa} contributes as 
 \be \la{FirstSa}
 \langle O \rangle \propto  \exp \left[ { m ( - \ell_{\rm hor} + R\gamma_0  - i R\chi_0 
  ) } \right]  = \exp \left[ { m ( - \ell_{\rm hor} + \ell_{\rm sing}   - i \tau_{\rm in}
  ) } \right]\,,
 \ee
 where $\ell_{\rm hor}$ is the regularized distance from the boundary to the horizon. 
 In order to understand the meaning of a positive value of $\gamma_0$, it is convenient to consider the low-temperature limit, where $r_- \sim r_+$. In this case,  both $\ell_{\rm hor}$ and $\gamma_0$ diverge because the distance to the horizon goes to infinity. However, this divergence cancels out in \nref{FirstSa}. This cancellation is obvious if we note that in this limit, $1/\sqrt{f}$ develops a pole at $u=u_+=u_-$, which the ${\cal C}_0$ contour goes around, see figure \ref{NearExt}. In the extremal limit, this contribution is not suppressed and becomes temperature independent. 
We interpret this as saying that this corresponds to the contributions of the Weyl tensor in the neck region --- the region that connects the AdS$_5$ geometry to the AdS$_2$ region\footnote{ We suggest this interpretation by noticing that,  in the low-temperature limit, we expect an almost decoupling of the physics of the $AdS_2$ region from the rest. Furthermore, we do not expect a one-point function in the strict $AdS_2$ limit. The fact that we still receive a finite contribution in this limit suggests to us that it is related to the connecting region. }.
Nevertheless, note that it still contains the time $\tau_{\rm in}$ between the inner and outer horizons! Notice that in \nref{FirstSa} the distances $\ell_{\rm hor}$ and $\ell_{\rm sing}$ are being {\it subtracted}. Therefore, when we look at the geodesics in figure \ref{Penrose}, we should not add the proper lengths of the spacelike sections, but subtract them.  
 
 Integrating along the contour ${\cal C}_1$ in figure \ref{NearExt}, we get a possible saddle-point value $\hat \rho_1$. Compared to $\hat \rho_0$, it contains an additional imaginary part 
    \be 
 \hat \rho_1 = \gamma_0 - i (\pi - \chi_0)\,,
 \ee 
 where we have used that the integral over a full large circle gives $-i \pi$. 
 The contours ${\cal C}_2$ and ${\cal C}_3$ in figure \ref{NearExt} give us 
 \be 
  \hat \rho_2 = - \gamma_0 - i ( \pi - \chi_0) 
 ~,~~~~~
  \hat \rho_3 = -\gamma_0 - i \chi_0 \,.
 \ee 
These look similar to the previous ones, except that the quantity $\gamma_0$ appears with the opposite sign. 
 Such contributions would lead to very suppressed terms at low temperatures (where $\gamma_0 \to \infty$), since we do not have the cancellations mentioned in \nref{FirstSa}. More precisely, $\rho_3$ gives a term of the form
   \be \la{Finre}
  \langle O \rangle ~~ \supset ~~ T^{ 2 \Delta'} e^{ -i \chi_0 \Delta} = T^{ 2 \Delta'} e^{ -i \pi \Delta'}\,,
  \ee 
  where  $\Delta'$ is the dimension of the field in the AdS$_2$ region, given by 
  \be 
  \Delta' = m R_{{\rm AdS}_2} = \Delta \left({ r_+ \over 2  \sqrt{ 1 + 2 r_+^2} }\right) ~,~~~~~{\rm for} ~~~ r_+ - r_-\to 0\,.
  \ee 
  We see that  the time between the inner and outer horizons is  $\pi R_{{\rm AdS}_2}$.
  
Notice that, in contrast with the black brane case, the AdS$_2$ limit yields a factor of $T^{2\Delta'}$ (as opposed to $T^{\Delta'}$). This is connected to the fact that, in the limit of perfect SL$(2)$ symmetry of AdS$_2$, the one-point functions are zero. We present a simple toy model of one-point functions in nearly-AdS$_2$ in appendix \ref{AppAdSTwo}.

\section{Conclusions and discussion }

\subsection{Summary } 

In this paper, we have proposed that the time to the singularity is contained in the thermal one-point functions. This information is extracted by analyzing the dependence on the mass, with the assumption that the higher-derivative coupling depends only on a power of the mass. This assumption is true in string theory.

For large mass, we have argued that we can perform a saddle-point analysis in terms of geodesics. Then we pointed out that there is a saddle point near the singularity once we assume a natural coupling between the massive particle and two gravitons. This is not a proof, as we did not show that this saddle point always contributes or that it is dominant. To present evidence for the contributions of this saddle we did the following. 

We analytically computed the thermal one-point function for a black brane and checked that the proposal is correct in this particular case. Furthermore, we gave a more detailed contour rotation argument that explains why the saddle point contributes, despite the fact that it does not lie on the original integration contour. We suspect that a similar argument can be made for other black holes, but we did not present a general rigorous argument. 
 
 For more general black holes, we examined the form of various possible saddle-point contributions, picking out the ones that we expect to contribute. For Schwarzschild-like black holes containing a spacelike singularity, we found that the saddle point that gives \nref{MainEq} is the dominant one when the imaginary part of the mass is sufficiently large. However, there can be larger contributions when the imaginary part of the mass is small. 
 
For black holes with an inner horizon, the structure of the answer is a bit different, see \nref{FirstSa}. The dominant contribution looks roughly like a geodesic that goes through the outer horizon, to the inner horizon, and then to the singularity, see figure \ref{Penrose}. The timelike region produces a ``phase'' proportional to the time between the outer and inner horizons, $\tau_{\rm in}$. The spacelike regions give contributions with opposite signs. This cancellation implies that this contribution becomes temperature independent in the extremal limit.  For this reason,  from the point of view of the original integral,  we can interpret this as a contribution from the region that connects the AdS$_2$ space to the higher-dimensional background. There are other subleading saddles which display a temperature dependence of the form 
 $T^{ 2 \Delta'}$ as we approach the extremal limit. These are expected to be contributions from the nearly AdS$_2$ region. 
 
It is interesting that these one-point functions can be computed in the bulk using the Euclidean black hole through an integral involving only the exterior. It is only the saddle-point approximation that brings in the interior. Note that the actual saddle is at some complex value of the radial position. It is only because this value is very close to the singularity that we can relate it to a property of the Lorentzian black hole. 

Of course, the interior of a collapsing black hole can be much more complicated and we wonder if any of the considerations here can be extended to that case. 

The considerations of this paper give us some very indirect access to the interior. Notice that this time to the singularity is a property of the thermal state and is independent of possible Lorentzian processes happening behind the horizon. For example, we can start from the two-sided black hole and send a shock wave at very early time on the left-hand side so that it sits just behind the future horizon of the right-hand side observer. The expectation values of the right-hand side observer are unchanged. However, the real Lorentzian time to the singularity, the one experienced by an observer falling through the shock wave, will change. 

\subsection{Three-dimensional case} 

When the bulk has three dimensions, there are no gravitons and no Weyl tensor. Furthermore, in the case of an infinite black string, the one-point functions are zero due to conformal symmetry. However, non-zero  one-point functions do arise for a finite area BTZ black hole \cite{Kraus:2016nwo}. These can be interpreted as arising from a three-point coupling between the field in question and the square of another field, with this other particle forming a loop around the black hole horizon. It would be interesting to see whether this mechanism also leads to \nref{MainEq}.  Naively, the same logic that leads to \nref{MainEq} should lead to a similar result for the three-dimensional case when we evaluate it perturbatively. We simply replace $W^2$ by the part of the loop diagram in the bulk that wraps non-trivially along the horizon. Here, we would also expect to obtain \nref{MainEq}, since diagram will get very large near the singularity. However, we could not see this formula from the analysis in \cite{Kraus:2016nwo}.  In appendix \ref{AdSthree}, we check that that in the BTZ case the one point function also contains the time to the singularity, at least in a particular case.

 \subsection{Two-point functions and thermal one-point functions of higher-spin operators}
 
Thermal one-point functions are relevant when we make an operator product expansion of two-point functions in a thermal background \cite{ElShowk:2011ag,Kulaxizi:2018dxo,Fitzpatrick:2019zqz,Li:2019zba,Fitzpatrick:2020yjb,Iliesiu:2018fao}. The reason that the thermal two-point function is different than the vacuum two-point function is the fact that operators that appear in the OPE acquire non-zero expectation values in the thermal state. These expectation values can in principle be related to vacuum OPE data \cite{ElShowk:2011ag,Iliesiu:2018fao}. 
 
In the case of free theories, we have that an operator creates particles, and these particles propagate fairly independently from each other. This is related to the observation that there is a significant contribution from higher-spin operators in the OPE, and that furthermore, these operators acquire a vacuum expectation value in the thermal background. In contrast, for theories with an Einstein gravity dual, the OPE in the thermal state has contributions only from multi-graviton states \cite{ElShowk:2011ag,Kulaxizi:2018dxo,Fitzpatrick:2019zqz,Li:2019zba,Fitzpatrick:2020yjb,Iliesiu:2018fao}. This is associated with the fact that the bulk particle feels it is moving in a gravitational background as a single particle, sometimes at speeds less than the boundary light speed, see \cite{Dodelson:2020lal} for a recent discussion.
   
   One could then be curious about the fate of the higher-spin operators as we increase the coupling of the boundary theory.  We know that they acquire a large anomalous dimension, which makes them look like massive particles in the bulk \cite{Gubser:1998bc,Witten:1998qj}.   Nevertheless, we still expect them to develop expectation values in the thermal state. Our discussion explains the origin of these expectation values. They are absent in the Einstein gravity approximation, but they appear once we include the $\alpha'$ corrections, even in the planar theory. These involve couplings between the higher-spin fields and two or more gravitons. Such higher-derivative corrections are present since these massive string states can decay into  gravitons. These then lead to one-point expectation values that can be estimated using the methods of this paper.  This gives a pleasing continuity to the description: the higher-spin operators are always present, and with non-zero thermal expectation values, but their contributions are suppressed when the boundary theory is strongly coupled.
     Notice that these expectation values for higher-spin operators are already present in the classical theory. In other words, in the normalizations of \nref{LagraSta}, the expectation value of $\varphi$ is of order one in $G_{\rm N}$, or the $1/N^2$ expansion, but they are suppressed by the gravity limit of small $\alpha'/R^2$.

~\\
\textbf{Acknowledgments}

JM is grateful to Akash Goel for some initial collaboration on this subject. 
We are also grateful to A. Almheiri, S. Giombi, A. Maloney, G. Mandal, F. Popov, D. Stanford, E. Witten, Y. Zhao for discussions.

JM was supported in part by U.S. Department of Energy grant DE-SC0009988. 
MG was supported in part by the Princeton University Department of Physics. 
% and the Simons Foundation grant 385600. 
%J.M. also thanks the  National Science Foundation under Grant No. NSF PHY-1748958, for the support of his work at the KITP. 

\appendix 
   
\section{Normalization of the correlators } 
\la{Normalization}

Here we discuss the normalization of the one-point function. We will use the extrapolate dictionary, defining the unnormalized correlators by taking limits of bulk correlators 
\be \widetilde O(x) = \lim_{z \to 0} \left[ z^{-\Delta} \phi(x,z) \right]\,,
\ee 
where the metric is 
%\begin{equation}
$ds^2=(d\vec{x}_d^{\,2}+dz^2)/z^2$,
%\end{equation}
and $\phi$ is a canonically normalized scalar field
\begin{align}
\label{AdS4action}
S&=\frac{1}{2}\int{\sqrt{g} \bigg((\nabla\phi)^2+m^2\phi^2\bigg)}\,.
\end{align}
%\subsection{Two-point function}
%\subsection{Two-point normalization}
%With the unnormalized one-point function computed, we must now use the two-point function in order to determine the appropriate normalization coefficient. We choose to unit-normalize the two-point functions. To do so, we no longer assume $\vec{x}$ independence, and Fourier transform our sourceless Euclidean action \eqref{AdS4action}. Doing so, we obtain the bulk-to-bulk Green's function equation of motion,
After going to Fourier space in the $\vec x_d$ coordinates, the Green's function obeys the equation
\begin{equation}
\label{2ptGdiffeq}
\partial_z\left(\frac{1}{z^{d-1}}\partial_z\tilde{G}(z,z',k)\right)-\left(\frac{k^2z^2+m^2}{z^{d+1}}\right)\tilde{G}(z,z',k)
=\delta(z-z')\, .
\end{equation}
The homogeneous solutions take the form of two Bessel functions: $\eta_1(z,k)=z^{d/2}I_\n(kz)$
and $\eta_2(z,k)=z^{d/2}K_\n(kz)$, where the index is $\n=\Delta-d/2$. The solution of \eqref{2ptGdiffeq} can then be written as
%With this, our Fourier transformed Green's function can be constructed to satisfy \eqref{2ptGdiffeq}:
\begin{equation}
\label{2ptG}
\tilde{G}(z,z',k)=-\big[\eta_1(z,k)\eta_2(z',k)\theta(z'-z)+\eta_1(z',k)\eta_2(z,k)\theta(z-z')\big]\, .
\end{equation}
%Now, since we are interested in the boundary two-point function, we must extract the small $z$ and $z'$ behavior\footnote{The approach here is essentially to take the bulk-to-bulk propagator, $G(z,x,z',x')$, and send $z$ and $z'$ to the AdS boundary at $z=0$, in order to obtain the boundary two-point function.} of the Green's function \eqref{2ptG} while maintaining the requirement of vanishing for $z$ and $z'$ at $\infty$. Doing so, we find that the boundary-limit behavior of the Green's function is given by,
For small $z$ and $z'$, this behaves like
\begin{equation}
\tilde{G}(z,z',k)
\sim
2^{d-1-2\Delta}k^{2\Delta-d}\left(\frac{\G(\frac{d}{2}-\Delta)}{\G(\Delta-\frac{d}{2}+1)}\right)(zz')^\Delta\,,
\end{equation}
 or,  Fourier transforming back into position space
%in order to obtain the solution in our original coordinates, giving,
\begin{equation}
\label{2ptEM}
\langle \widetilde O(x) \widetilde O(0) \rangle = \lim_{ z,z' \to 0} ( z z')^{-\Delta }G(z,z',\vec{x})=
\frac{1}{2\pi^{d/2}}
\frac{\G(\Delta)}{\G(\Delta-\frac{d}{2}+1)}
\frac{1}{|x|^{2\Delta}}\,.
\end{equation}
We then conclude that the properly normalized operator is defined as 
\be \la{CNdef}
{\cal O}(x) = C_{\rm N}  \tilde O(x) = C_{\rm N} \lim_{z\to 0} \left[ z^{-\Delta} \, \phi(x,z) \right] ~,~~~~~C_{\rm N} \equiv \sqrt{ 2 \pi^{d/2} \, \Gamma(\Delta - { d \over 2 }+1 )
\over \Gamma(\Delta) }  \,.
\ee 
 This formula gives us the normalized one-point function. 
 Starting from 
 \be 
 \langle \phi(z',0) \rangle = \int dz dx \, \sqrt{g}\, G(z',0;z,x) \, \hat \alpha W^2~,
 \ee 
 the boundary expectation value is obtained by extracting the ${z'}^{\Delta}$ piece of this expectation value. 
We can take the $z'\to 0$ limit first inside the integral, and then use the $z'< z$ form of the propagator \nref{propaBB} to obtain \nref{IntRe}. 
The factors of $(16 \pi G_{\rm N})$ arise from the normalization of \nref{LagraSta}. Similarly, the factors of $R$ can be easily restored. 
 
 \subsection{Normalization in the geodesic approximation } 
\la{GravNor} 

A simple way to determine the normalization in the geodesic approximation is the following. The unit normalization at short distances, $ \langle O(x) O(0) \rangle \sim |x|^{ -2\Delta}$, implies that operators on the sphere behave like, $\langle O(\theta) O(0) \rangle \sim [ 2 \sin { \theta \over 2 } ]^{ -2 \Delta}$. On opposite points, we then have  $\langle O(\pi) O(0) \rangle \sim 2^{ - 2 \Delta } $. 

Writing the empty AdS metric as in \nref{SchAdSd} with $f(r) = r^2 +1$, we find that the two-point function is given by $e^{-\Delta \ell }$. Here $\ell$ is the total length, given explicitly by
\be 
\ell = 2 \int_0^{r_{\rm c}} { dr \over \sqrt{r^2 +1} } =  2 \log r_{\rm c} + 2 \log( 2) \, .
\ee
This implies that we simply need to subtract a factor of $\log r_{\rm c}$ for each operator, with no extra constant. Then the $\log(2)$ term correctly reproduces the expected answer. Of course, we get the same prescription if we use the usual semicircular geodesics in Poincar\'e coordinates. 

\section{Prefactor } 
\la{Prefactor}

With translation symmetry, the propagator obeys the wave equation 
\be 
 { 1 \over r^{d-1} } \partial_r ( f(r) r^{d-1} \partial_r \Psi) - m^2 \Psi =0 \, ,
 \ee 
 away from coincident points. 
 The standard WKB method then gives solutions 
 \be 
 \Psi \sim F(r)  \exp\left( \pm m\int^r {dr'\over \sqrt {f(r')}}  \right) ~,~~~~F(r) \equiv { 1 \over r^{d-1 \over 2} \sqrt{f(r)} }\, .
\ee 
The prefactor $F$ gives rise to additional singularities at positions where $f=0$.

For the particular case of the singularity at the horizon, we can choose a new variable, $\rho$  \nref{PropDi}. The equation near $\rho=0$ is just that of the Bessel function, since the cigar looks like two-dimensional Euclidean space. The regular solution is simply the $I_0(m\rho)$ function, which can be expanded for large mass as 
\be \la{Bess}
I_0 \propto  { e^{ m \rho } \over \sqrt{ m \rho } }(1 + \cdots) - i { e^{ - m \rho } \over \sqrt{ m \rho } }(1 + \cdots )~,~~~~~{\rm for}~~{\rm Im}(m\rho) < 0\,,
\ee 
where each term is also multiplied by powers of $(m\rho)^{-1}$.  
This gives us the relative normalization of the short and long geodesic contributions near $\rho=0$. The fact that there is an $i$ for the long geodesic is reasonable because it is expected to have a negative mode. The long geodesic integral involves the second factor in \nref{Bess}
\be \la{ILong}
I_{\rm long} = -i \int_0^\infty d\rho  \rho   { e^{ - m\rho } \over \sqrt{\rho} }  \, ,
\ee 
where we have indicated only the small-$\rho$ behavior and neglected unimportant overall factors. The factor of $\rho$ comes from the volume of the circle. 
The short geodesic contribution involves the first term in \nref{Bess}. Integrating along the contour ${\cal C}_{\rm s-left}$ in figure \ref{ContPlanar}, we find
\be \la{IShort}
I_{\rm short} = \int_0^{-\infty} d\rho \rho ~  { e^{ m \rho} \over \sqrt{  \rho } } =  i\int_{0}^{\infty} d\rho' \rho'   { e^{ - m \rho' } \over \sqrt{\rho'} } ~,~~~~~\rho = e^{ - i \pi } \rho' \, ,
\ee 
which cancels \nref{ILong}. There is a similar cancellation if we use the full prefactor for the black brane
\be \la{BBPre}
F(\rho) \propto { 1 \over \sqrt{ \sinh \rho } } \, ,
\ee
which replaces the $1/\sqrt{\rho}$ in \nref{Bess}.
This cancellation is important for the result we are obtaining. If we had not had this cancellation, each integral would have only given powers of $m$ and would have been larger than the term going like $e^{ -i \pi \Delta/d}$, which is very small for Im($\Delta)<0$. 

The prefactor \nref{BBPre} also has singularities at $\rho = -i n$, which seem to interfere with our contour rotation argument. We have not understood how to treat these properly. Perhaps one should consider the saddle-point approximation in the two-dimensional space of $\rho$ and $t_{\rm E}$, where $t_{\rm E}$ is the Euclidean time direction, after a suitable complexification. This should be doable in terms of Lefschetz thimbles, see \cite{Witten:2010cx}. In the unlikely case that they do not cancel, they would give contributions involving exponentials of  $\rho = -i\pi n$. The first coincides with the leading contribution we have kept. And the others would be subleading 
 if Im$(m) < 0$, so that they would not affect \nref{MainEq}. 

\section{Geodesic approximation for three-point functions} 
\la{ThreePoints}

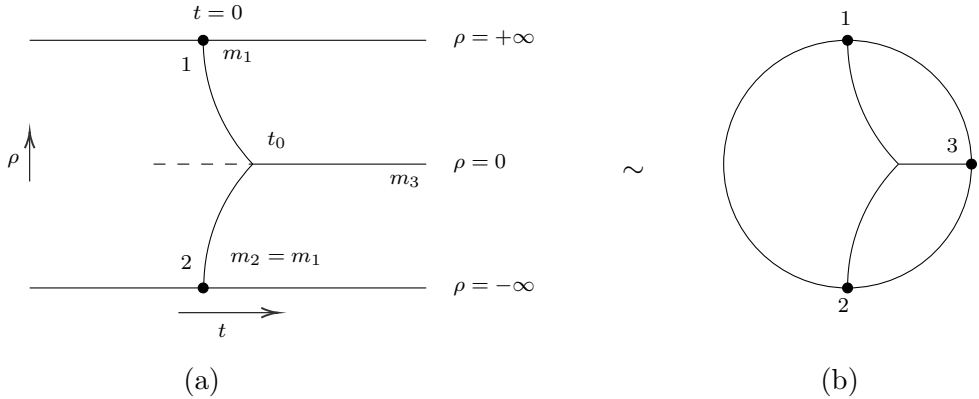
\begin{figure}[h]
\begin{center}
\begin{tikzpicture}[x=0.75pt,y=0.75pt,yscale=-1.25,xscale=1.25]
\draw  [dash pattern={on 4.5pt off 4.5pt}]  (170,90) -- (210,90) ;
\draw    (210,90) -- (280,90) ;
\draw   (190,140) .. controls (190.26,139.52) and (190.26,139.45) .. (190.26,139.37) .. controls (190.26,120.01) and (198.12,102.49) .. (210,90) ;
\draw   (210,90) .. controls (197.99,77.15) and (190.03,59.56) .. (190,40) ;
\draw   (450,140) .. controls (450,139.93) and (450,139.85) .. (450,139.78) .. controls (450,120.42) and (457.86,102.89) .. (470.56,90) ;
\draw    (470.56,90) -- (500,90) ;
\draw   (470.56,90) .. controls (457.99,77.15) and (450.03,59.56) .. (450,40) ;
\draw    (120,40) -- (280,40) ; 
\draw    (120,140) -- (280,140) ; 
\draw   (400,90) .. controls (400,62.39) and (422.39,40) .. (450,40) .. controls (477.61,40) and (500,62.39) .. (500,90) .. controls (500,117.61) and (477.61,140) .. (450,140) .. controls (422.39,140) and (400,117.61) .. (400,90) -- cycle ;
\draw    (180,150) -- (218,150) ;
\draw [shift={(220,150)}, rotate = 180] [color=black][line width=0.75]    (6.56,-1.97) .. controls (4.17,-0.84) and (1.99,-0.18) .. (0,0) .. controls (1.99,0.18) and (4.17,0.84) .. (6.56,1.97)   ; 
\draw    (120,77) -- (120,97) ;
\draw [shift={(120,77)}, rotate = 450] [color=black][line width=0.75]    (6.56,-1.97) .. controls (4.17,-0.84) and (1.99,-0.18) .. (0,0) .. controls (1.99,0.18) and (4.17,0.84) .. (6.56,1.97)   ;
\draw (181,171) node [anchor=north west][inner sep=0.75pt]   [align=left] {(a)};
\draw (438,171) node [anchor=north west][inner sep=0.75pt]   [align=left] {(b)};
\draw (195,153.4) node [anchor=north west][inner sep=0.75pt]  [font=\scriptsize]  {$t$};
\draw (110,85) node [anchor=north west][inner sep=0.75pt]  [font=\scriptsize]  {$\rho $};
\draw (215,76) node [anchor=north west][inner sep=0.75pt]  [font=\scriptsize]  {$t_{0}$};
\draw (264.27,93.4) node [anchor=north west][inner sep=0.75pt]  [font=\scriptsize]  {$m_{3}$};
\draw (200,125) node [anchor=north west][inner sep=0.75pt]  [font=\scriptsize]  {$m_{2} =m_{1}$};
\draw (196.96,42.87) node [anchor=north west][inner sep=0.75pt]  [font=\scriptsize]  {$m_{1}$};
\draw (290,35) node [anchor=north west][inner sep=0.75pt]  [font=\scriptsize]  {$\rho =+\infty $};
\draw (290,85) node [anchor=north west][inner sep=0.75pt]  [font=\scriptsize]  {$\rho =0$};
\draw (290,135) node [anchor=north west][inner sep=0.75pt]  [font=\scriptsize]  {$\rho =-\infty $};
\draw (358,90) node [anchor=north west][inner sep=0.75pt]   [align=left] {$\sim$};
\draw (446,27.4) node [anchor=north west][inner sep=0.75pt]  [font=\scriptsize]  {$1$};
\draw (489,78.4) node [anchor=north west][inner sep=0.75pt]  [font=\scriptsize]  {$3$};
\draw (445,143.4) node [anchor=north west][inner sep=0.75pt]  [font=\scriptsize]  {$2$};
\draw (180,46) node [anchor=north west][inner sep=0.75pt]  [font=\scriptsize]  {$1$};
\draw (180,126) node [anchor=north west][inner sep=0.75pt]  [font=\scriptsize]  {$2$};
\draw (185,25) node [anchor=north west][inner sep=0.75pt]  [font=\scriptsize]  {$t=0$};
\filldraw (500,90) circle (1.5pt);
\filldraw (450,140) circle (1.5pt);
\filldraw (450,40) circle (1.5pt);
\filldraw (190,140) circle (1.5pt);
\filldraw (190,40) circle (1.5pt);
\end{tikzpicture}
\caption{Geodesics for the three-point function. (a) The coordinates used in \nref{DiskMe}. Configuration for the case $m_1=m_2 > m_3/2$. (b) Conventional picture in terms of the hyperbolic disk.}
\label{ThreePoint}
\end{center}
\end{figure}

As shown in \cite{Bargheer_2014},  the Witten diagram for the three-point function for large masses can be approximated by a geodesic computation, as shown in figure \ref{ThreePoint}(b). This reproduces the large-$\Delta_i$ limit of the gamma functions appearing in the Witten diagram computed in  \cite{Freedman:1998tz}. 
The approximation involves writing each propagator in terms of geodesics and using a saddle point for the integration over the interaction point.  A real saddle point exists if the masses obey $m_1 + m_2 > m_3$ (up to permutations). 

However, if  $m_1 + m_2 < m_3$, then the interaction point gets driven to the boundary, and more specifically, to the insertion point of the third operator. This is related to the appearance of poles at $\Delta_3 = \Delta_1 + \Delta_2 + 2 n$, which stem from the mixing of the $O_3$ with operators of the schematic form $O_1 \partial ^{ 2 n} O_2$. 
Nevertheless, even in this case, it is possible to show that  we can reproduce the large-mass (large-$\Delta$) behavior from a complex solution.    

Since we just want to illustrate the phenomenon, we will choose a simple case with $m_1 = m_2 \not = m_3$. We can consider all three points at the boundary of an $H_2$ bulk subspace with coordinates 
\be 
\la{DiskMe}
ds^2 = d\rho^2 + \cosh^2 \rho \, dt^2\,.
\ee 
We put $O_1$ and $O_2$ at $t=0 $ and $\rho = \pm \infty$. We also place the third operator at $\rho =0 $ and $t= + \infty$, see figure \ref{ThreePoint}. By symmetry, the classical trajectory of the third particle is at $\rho=0$. The first particle follows a trajectory 
\be 
 \tanh \rho =   \cosh t - {  \sinh t \over \tanh t_0 } ~,~~~~~~  0 \leq t \leq t_0\,,
\ee
where $t_0$ is the value of $t$ at $\rho=0$ (the intersection point with the third particle), see figure \ref{ThreePoint}.  The second particle is at a symmetric configuration. 

Evaluating its length, we get 
 \be
 \ell(t_0) = \int_{\epsilon}^{t_0} dt \sqrt{ \cosh^2 \rho + \rho'^2 } = \half \log \left({ \sinh 2 t_0 \over \epsilon }\right) = \log \cosh t_0  + \rho_{\rm max} \, ,
 \ee
where we have used the relation $\rho_{\rm max} \sim  \half (- \log \epsilon +  \log \tanh t_0 +\log 2 ) $, and $\rho_{\rm max}$ is a physical cutoff, independent of $t_0$. The final action is then
 \be \la{GeoAct}
 S = R \, \left[ 2 m_1 \log \cosh{t_0} - m_3 t_0 \right] \, .
\ee
Minimizing with respect to $t_0$, we find
\be 
\tanh t_0 = { m_3 \over 2 m_1}
~,~~~~~{\rm or}~~
\tanh u_0 = { 2 m_1 \over m_3 }
~,~~~~{\rm for }~~
t_0 = u_0 + i \pi/2 \,.
 \ee
 We see that $t_0$ is complex  for $m_3> 2 m_1$. 
 %But there is a complex solution with 
 %\be 
 %t_0 = u_0 + i \pi/2 ~,~~~~~~~ \tanh u_0 = { 2 m_1 \over m_3 } 
% \ee
 The action \nref{GeoAct} becomes 
 \be \la{FinStp}
 S =  i \pi( 2 \Delta_1 -\Delta_3)/2  + \left({\rm real} \right)\, .
 \ee
 This reproduces the ``phase'', $e^{ i \pi (2 \Delta-1-\Delta_3)/2}$ of the 
% which when we replace the parenthesis by $ 2n$ gives an expected oscillatory sign each time we cross a pole of the factor 
$\Gamma\left((\Delta_1 + \Delta_2 -\Delta_3)/2\right)$ factor in the Witten diagram.
% In other words, this factor is related to the ``phase'' 
 % $e^{ i \pi (2 \Delta-1-\Delta_3)/2}$. 
 By summing over saddles with  $t_0 \to t_0 + i \pi n$, we reproduce a $1/\sin(\pi(2\Delta-\Delta_3)/2)$ factor present from the gamma function. The real part in \nref{FinStp} reproduces the large-$\Delta_i$ limit of all other gamma factors.   

The conclusion is that in this well-studied example, we also find that complex saddle points reproduce the answer.

 \section{Toy model for one-point functions in nearly-AdS$_2$} 
 \la{AppAdSTwo}
 
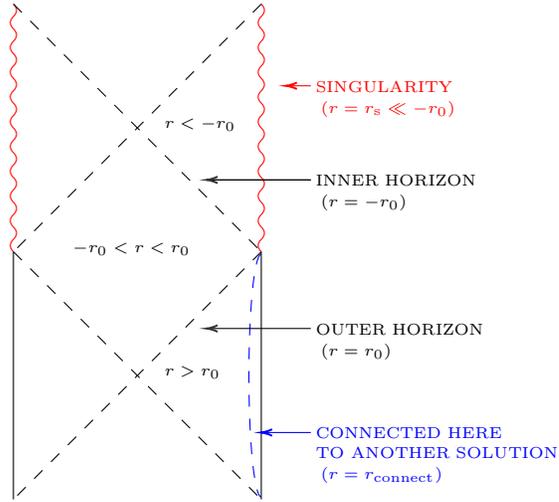
\begin{figure}[h]
\begin{center}
\makebox{
\hspace{38mm}
\begin{tikzpicture}[x=0.75pt,y=0.75pt,yscale=-1.25,xscale=1.25]
\draw [color=blue, draw opacity=1, dash pattern={on 4.5pt off 4.5pt}] (380,111) arc (270:90:5 and 49);
\draw [color=red  ,draw opacity=1 ]   (280,10) .. controls (281.67,11.67) and (281.67,13.33) .. (280,15) .. controls (278.33,16.67) and (278.33,18.33) .. (280,20) .. controls (281.67,21.67) and (281.67,23.33) .. (280,25) .. controls (278.33,26.67) and (278.33,28.33) .. (280,30) .. controls (281.67,31.67) and (281.67,33.33) .. (280,35) .. controls (278.33,36.67) and (278.33,38.33) .. (280,40) .. controls (281.67,41.67) and (281.67,43.33) .. (280,45) .. controls (278.33,46.67) and (278.33,48.33) .. (280,50) .. controls (281.67,51.67) and (281.67,53.33) .. (280,55) .. controls (278.33,56.67) and (278.33,58.33) .. (280,60) .. controls (281.67,61.67) and (281.67,63.33) .. (280,65) .. controls (278.33,66.67) and (278.33,68.33) .. (280,70) .. controls (281.67,71.67) and (281.67,73.33) .. (280,75) .. controls (278.33,76.67) and (278.33,78.33) .. (280,80) .. controls (281.67,81.67) and (281.67,83.33) .. (280,85) .. controls (278.33,86.67) and (278.33,88.33) .. (280,90) .. controls (281.67,91.67) and (281.67,93.33) .. (280,95) .. controls (278.33,96.67) and (278.33,98.33) .. (280,100) .. controls (281.67,101.67) and (281.67,103.33) .. (280,105) .. controls (278.33,106.67) and (278.33,108.33) .. (280,110) -- (280,110) ;
\draw [color=red  ,draw opacity=1 ]   (380,10) .. controls (381.67,11.67) and (381.67,13.33) .. (380,15) .. controls (378.33,16.67) and (378.33,18.33) .. (380,20) .. controls (381.67,21.67) and (381.67,23.33) .. (380,25) .. controls (378.33,26.67) and (378.33,28.33) .. (380,30) .. controls (381.67,31.67) and (381.67,33.33) .. (380,35) .. controls (378.33,36.67) and (378.33,38.33) .. (380,40) .. controls (381.67,41.67) and (381.67,43.33) .. (380,45) .. controls (378.33,46.67) and (378.33,48.33) .. (380,50) .. controls (381.67,51.67) and (381.67,53.33) .. (380,55) .. controls (378.33,56.67) and (378.33,58.33) .. (380,60) .. controls (381.67,61.67) and (381.67,63.33) .. (380,65) .. controls (378.33,66.67) and (378.33,68.33) .. (380,70) .. controls (381.67,71.67) and (381.67,73.33) .. (380,75) .. controls (378.33,76.67) and (378.33,78.33) .. (380,80) .. controls (381.67,81.67) and (381.67,83.33) .. (380,85) .. controls (378.33,86.67) and (378.33,88.33) .. (380,90) .. controls (381.67,91.67) and (381.67,93.33) .. (380,95) .. controls (378.33,96.67) and (378.33,98.33) .. (380,100) .. controls (381.67,101.67) and (381.67,103.33) .. (380,105) .. controls (378.33,106.67) and (378.33,108.33) .. (380,110) -- (380,110) ;
\draw  [dash pattern={on 4.5pt off 4.5pt}]  (380,10) -- (280,110) ;
\draw  [dash pattern={on 4.5pt off 4.5pt}]  (280,10) -- (380,110) ;
\draw  [dash pattern={on 4.5pt off 4.5pt}]  (380,110) -- (280,210) ;
\draw  [dash pattern={on 4.5pt off 4.5pt}]  (280,110) -- (380,210) ;
\draw  (280, 110) -- (280,210); 
\draw  (380, 110) -- (380,210); 
\draw  [color = blue]  (400,183) -- (380,183) ;
\draw [shift={(380,183)}, rotate = 0] [color=blue  ][line width=0.75]    (4.37,-1.32) .. controls (2.78,-0.56) and (1.32,-0.12) .. (0,0) .. controls (1.32,0.12) and (2.78,0.56) .. (4.37,1.32)   ;
\draw    (400,81) -- (358,81) ;
\draw  [color = red]  (400,43) -- (390,43) ;
\draw [shift={(390,43)}, rotate = 0] [color=red  ][line width=0.75]    (4.37,-1.32) .. controls (2.78,-0.56) and (1.32,-0.12) .. (0,0) .. controls (1.32,0.12) and (2.78,0.56) .. (4.37,1.32)   ;
\draw    (400,81) -- (358,81) ;
\draw [shift={(358,81)}, rotate = 0] [color=black  ][line width=0.75]    (4.37,-1.32) .. controls (2.78,-0.56) and (1.32,-0.12) .. (0,0) .. controls (1.32,0.12) and (2.78,0.56) .. (4.37,1.32)   ;
\draw    (400,141) -- (358,141) ;
\draw [shift={(358,141)}, rotate = 0] [color=black  ][line width=0.75]    (4.37,-1.32) .. controls (2.78,-0.56) and (1.32,-0.12) .. (0,0) .. controls (1.32,0.12) and (2.78,0.56) .. (4.37,1.32)   ;
\draw (303,105) node [anchor=north west][inner sep=0.75pt]  [font=\footnotesize,opacity=1 ]  {\tiny $-r_0<r<r_0$};
\draw (340,155) node [anchor=north west][inner sep=0.75pt]  [font=\footnotesize,opacity=1 ]  {\tiny $r>r_0$};
\draw (340,55) node [anchor=north west][inner sep=0.75pt]  [font=\footnotesize,opacity=1 ]  {\tiny $r<-r_0$};
\draw (401,40) node [anchor=north west][inner sep=0.75pt]  [font=\tiny,color=red ,opacity=1 ] [align=left] {\textcolor{red}{SINGULARITY}\\};
\draw (403,48) node [anchor=north west][inner sep=0.75pt]  [font=\tiny,color=red ,opacity=1 ] [align=left] {\textcolor{red}{$(r=r_{\rm s}\ll -r_0)$}\\};
\draw (401,78) node [anchor=north west][inner sep=0.75pt]  [font=\tiny] [align=left] {INNER HORIZON};
\draw (403,86) node [anchor=north west][inner sep=0.75pt]  [font=\tiny] [align=left] {$(r=-r_0)$};
\draw (401,138) node [anchor=north west][inner sep=0.75pt]  [font=\tiny] [align=left] {OUTER HORIZON};
\draw (403,146) node [anchor=north west][inner sep=0.75pt]  [font=\tiny] [align=left] {$(r=r_0)$};
\draw (401,180) node [anchor=north west][inner sep=0.75pt]  [font=\tiny,color=darkgreen ,opacity=1 ] [align=left] {\textcolor{blue}{CONNECTED HERE}\\};
\draw (401,188) node [anchor=north west][inner sep=0.75pt]  [font=\tiny,color=darkgreen ,opacity=1 ] [align=left] {\textcolor{blue}{TO ANOTHER SOLUTION}\\};
\draw (403,196) node [anchor=north west][inner sep=0.75pt]  [font=\tiny,color=darkgreen ,opacity=1 ] [align=left] {\textcolor{blue}{$(r=r_{\rm connect})$}\\};
\end{tikzpicture}}
\caption{Penrose diagram of nearly-AdS$_2$ spacetime. The exterior region, $r>r_0$, is then connected to some of the spacetime at $r=r_{\rm connect}$. The region beyond the inner horizon contains a timelike singularity at $r = r_s \ll -r_0$.   }
\label{AppPenrose}
\end{center}
\end{figure}

 We can consider the AdS$_2$ metric 
 \be \la{AdSm}
 ds^2 = R^2 \left[- (r^2-r_0^2) dt^2 + { dr^2 \over r^2 -r_0^2 } \right] \,.
 \ee
 In purely-AdS$_2$, any expectation value has to be a constant, which we can subtract. 
Notice that the region $ - r_0 < r < r_0$ corresponds to the region between the outer and inner horizons, see figure \ref{AppPenrose}. 
 
If we now consider a space which is nearly-AdS$_2$ (as it arises when we take the near-extremal limit of a more general black hole) then the metric \nref{AdSm} will connect to some other space at some large value, $r =r_{\rm connect}\gg r_0$. Similarly, we expect deviations in the   region behind the inner horizon $(r = r_s \ll -r_0)$, see figure \ref{AppPenrose}.
 
We expect that these deviations will induce some effective coupling to the scalar field of the form 
 \be 
 S_{\rm source} = \int dt_{\rm E} dr \,\sqrt{g} \, f(r) \, \phi(t_{\rm E},r)  ~,~~~~~~~~ 
 %\beta \propto 1/r_0 
 \ee
 where $f(r)$ is some function. The integral is over the Euclidean black hole, which contains only the exterior region $r\geq r_0$. As a toy model, we choose the function 
 \be \la{fExp}
 f(r) = \left[ { -r_{\rm s} \over r - r_{\rm s}} \right]^2 \,,
 \ee
 which has some desirable features. First, it goes to zero at the physical boundary, $(r \to + \infty)$. It also diverges at $r = r_s \ll 0$, which is in the region where we expect the singularity to lie. We keep $r_s$ fixed as we vary the temperature, or vary $r_0\propto T$. We note that the particular function \nref{fExp} was only chosen so that we can analytically compute the integrals below. 
 
 We can solve for the propagator, as in the black brane case, after choosing the variable $w \equiv { 2 r_0 \over r +r_0 }$.
However here, in contrast with section \ref{AnBB}, $w$ continues beyond infinity, to negative values, where it describes the region behind the inner horizon $(r< -r_0)$. see figure \ref{AppPenrose}. 
 The expression for the one-point function then becomes, (assuming $r_0/|r_{\rm s}| \ll 1$)
 \be 
 \langle O \rangle \propto r_0^{\Delta} \int_{0}^1  { d w \over w^2 }  \left[ {w \over w - w_{\rm s} } \right]^2 w^{\Delta} ~_2F_1(\Delta,\Delta,1; 1-w )   ~,~~~ w_{\rm s} = { 2 r_0 \over r_{\rm s}} <0 ~,~~~\Delta \sim  m R \,,
 \ee 
where the integral is only over the black hole exterior. This gives 
  \bea \la{ExRes}
   \langle O \rangle & \propto &  r_0^\Delta { \Gamma( \Delta +1) \Gamma(2 -\Delta )   } { 1\over w_{\rm s}^2 } ~_2F_1\left( \Delta +1, 2-\Delta , 2 ;  {1 \over w_{\rm s} } \right)
   \cr  
& \propto &  {\Gamma(2- \Delta ) \Gamma(2\Delta  -1) \over \Gamma(\Delta )  }    \left( { - r_{\rm s} \over 2 }\right)^{ \Delta } + \cdots
{ \Gamma( \Delta  +1) \Gamma(1 - 2 \Delta  ) \over \Gamma(1-\Delta )   } r_0^{2 \Delta -1} \left( { - r_{\rm s} \over 2 }\right)^{ 1-\Delta  }  + \,\cdots~~~~~~~~~ 
    \eea 
    where the dots represent extra integer powers of $r_0$. 
Note that, for large $\Delta $, both terms lead to a ``phase'' factor of the form $e^{ - i \Delta  \pi }$, which comes from the large-$\Delta$ expansion of the gamma functions. The $ \pi R $ here is indeed the time between the inner and outer horizons for \nref{AdSm}. The first term in \nref{ExRes} is like the temperature-independent term that was discussed in \nref{FirstSa}. The second term gives the temperature-dependent term, as in \nref{Finre}, since $r_0 \propto T$. \footnote{In \nref{Finre}, the approximation does not distinguish between $2\Delta'$ and $2\Delta' -1$. $\Delta' $ in \nref{Finre} is the same as $\Delta $ here --- the scaling dimension in the AdS$_2$ region.}

Note that the temperature dependence of these two terms can also be obtained as follows. Suppose we have a perturbation
\be \la{PertAc}
S  = \eta \int dt O(t) \,
\ee
at the boundary. Assuming the conformally invariant two-point function 
\be 
 \langle O(t) O(0) \rangle = { T^{2 \Delta } \over (\sin (T \pi t) )^{2 \Delta}  }   \, ,
 \ee
perturbing the theory with \nref{PertAc}, gives 
 \be \la{IntTOO}
 \langle O \rangle \sim \eta \int dt \langle O(t) O(0) \rangle \propto T^{2 \Delta  -1 } \, ,
 \ee
  where the last term comes from a rescaling of the integration variables or dimensional analysis. This reproduces the temperature dependence of the second term in \nref{ExRes}. The first term, which is temperature independent, comes from the UV divergence of \nref{IntTOO}. This simple integral \nref{IntTOO} does not however, reproduce the ``phase'' factor that we encountered above, so the story is not complete.  

The additional integer powers of $T$ or $r_0$ present in \nref{ExRes}, can be viewed as arising from terms involving 
$\eta c_n \int dt O H^n $, where $H$ is the Hamiltonian\footnote{We thank the anonymous referee for this comment}.

\section{Thermal one point functions for three dimensional black holes}
\la{AdSthree}
 
\begin{figure}[h]
\begin{center}
\begin{tikzpicture}
[x=0.75pt,y=0.75pt,yscale=-1.25,xscale=1.25]
\draw (60,0) arc (0:360:60);
\filldraw (0,0) circle (10pt);
\draw [color=red] (20,0) arc  (0:360:20);
\draw [color=blue]   (20,0) -- (60,0);
\draw (20,-20) node [color=red,opacity=1] [align=left] {\textcolor{red}{$\chi$}};
\draw (67,0) node [color=blue,opacity=1] [align=left] {\textcolor{blue}{$\phi$}};
\end{tikzpicture}
\caption{Origin of the thermal one point function for a BTZ black hole \cite{Kraus:2016nwo}.}
\label{Threedcase}
\end{center}
\end{figure}
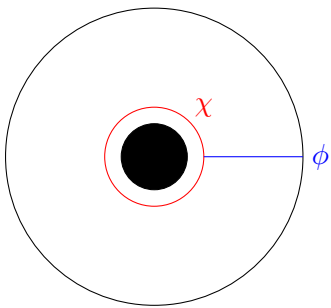

%BTZ $\leftrightarrow$ thermal AdS$_3$ (up to modular transformation)
Thermal one point functions are zero for an infinite black string due to conformal symmetry. However, as shown in 
 \cite{Kraus:2016nwo} they are non-zero once we compactify the spatial direction and obtain a  BTZ black hole.   As argued in \cite{Kraus:2016nwo},  the one-point function can be viewed as arising from a bulk diagram where the field in question, $\phi$,  interacts via a three-point coupling with another field, $\chi$,  that goes around the horizon of the BTZ black hole, see figure \ref{Threedcase}. As an example, consider an action of the form 
\begin{equation}
S \propto \int (\nabla\phi)^2+m^2\phi^2
+ (\nabla\chi)^2+\mu^2 \chi^2 + g \, \phi\chi^2 \,,
\end{equation}
where $\phi$ and $\chi$ are bulk scalar fields, and $\phi$ is related to the operator whose one-point function we are after. This one-point function comes from the interactions with the field $\chi$. More precisely, we get a contribution of the form  
\begin{equation} \la{Oxp}
\langle O\rangle \propto \int_{BTZ} \langle\chi^2(x)\rangle_\beta G_b(x,0)\,,
\end{equation}
where $\beta$ denotes temperature dependence. And $G_b$ is a bulk to boundary propagator.  In the coincident point correlator for the field $\chi$, $\langle \chi^2(x) \rangle$, we include only the contribution from trajectories with non-zero winding around the spatial circle, disregarding the contribution from the unwound trajectories which should be removed when we are in global AdS$_3$. In other words, when we go from global AdS$_3$ to the BTZ black hole, we perform a quotient. In computing the $\chi$ correlator we use the method of images, keeping only the contributions from non-trivial images. This means that $\langle \chi^2(x) \rangle$ is some function of the radial coordinate. From now on the analysis is similar to what we did in higher dimensions. More explicitly, we can write Euclidean  metric as 
\be \la{BTZ}
ds^2 =  { 1 \over z^2 } \left(   (1 -z^2) dt^2 + { dz^2 \over 1 -z^2} + z^2 dx^2 \right)  \,, ~~~~t \sim t + 2 \pi
\ee 
where we have rescaled the time coordinate to set the temperature to $2\pi$. 
The length of the $x$ coordinate is then
\be 
x \sim x + \ell ~,~~~~~~~~ \ell =   2\pi L/\beta \,,
\ee
where $\beta$ is the original temperature and $L$ is the original length of the circle. 

If we momentarily ignore the compactification of $x$, we can view \nref{BTZ} as global $AdS_3$ where the angular direction is $t$. The propagator of the $\chi$ field is given by \cite{DHoker:1999mqo}  
 \begin{equation} 
G_\De = \frac{C_\De 2^\De}{u^\De}
~_2F_1\left(\Delta,\Delta-\frac{1}{2} ,2\Delta-1;-\frac{2}{u}\right)
\ee
with
\be 
u \equiv  2 \sinh^2 {\hat d\over 2 }\,,  \la{Udef}
\end{equation}
where  $\hat d $ is the proper distance between the two points and $\Delta = 1 + \sqrt{ 1 + \mu^2}$ (we have set the $AdS_3$ radius to one).   Then we can write 
\begin{equation} \la{SumS}
\langle\chi^2(x)\rangle  
=\sum_{n=1}^\infty  G(u_n),
\end{equation}
where we are summing over images, but have subtracted the completely coincident point singularity.   
%
%Now, for AdS$_{d+1}$ propagators  we have
%\begin{equation}
%G_\De = \frac{C_\De 2^\De}{u^\De}
%F\left(\Delta,\Delta-\frac{d}{2}+\frac{1}{2},2\Delta-d+1;-\frac{2}{u}\right)
%~,~~~
%u=\frac{(z-z')^2+(\vec x-\vec x ')^2}{2zz'}\,.
%\end{equation}
We have that $u_n$ is given, in terms of the distance $\hat d_n$ from a general point to its image, by \nref{Udef},   \be \la{Dis}
u_n = 2 \sinh^2{ \hat d_n \over 2} ~,~~~~~\sinh {\hat d_n \over 2} = { \sinh ({ n \ell \over 2}) \over z }  \, .
\ee 
 We see from these expressions that when $z\to \infty$, $\hat d_n$ or $u_n \to 0$ and the terms in the sum \nref{SumS} diverge. In fact,  since this expression appears in \nref{Oxp} we therefore expect that there will be saddle point solutions as we had previously. We will now check this explicitly. 
 
Consider a particular dimension for the $\chi $ field, $\Delta =1$,   where the propagator \nref{Udef} becomes relatively simple
\begin{equation} \la{ChiP}
G\propto -\frac{2}{u}\,F\left(1,\half,1;-{2 \over u}\right) \propto \frac{1}{\sqrt{u} \sqrt{1+ u/2}} \propto { 1 \over \sinh \hat d} \, .
\end{equation}
Let us first insert the $n=1$ from \nref{SumS} in the expression \nref{Oxp}. 
 Defining $w =z^2$, we find 
\begin{equation}
\langle\mathcal{O}\rangle
\propto \int_0^1 { d z \over z^3}  \langle\chi^2(z)\rangle_{n=1} G_b(z)
%\propto \int_0^1 \frac{dz}{z^3}
%\frac{1}{\frac{(\cosh\tilde{\beta}-1)}{z^2}+1}z^{\Delta_\phi}
%F\left[\frac{\Delta_\phi}{2},\frac{\Delta_\phi}{2},1;(1-z^2)\right]\,.
%\end{equation}
%Now, let $h\equiv\Delta_\phi/2$ and $w=z^2$, so $\frac{dz}%{z^3}\propto\frac{dw}{w^2}$. Then we have
%\begin{equation}
%\langle\mathcal{O}\rangle=
\propto { 1 \over \sinh { \ell \over 2 } }\int_0^1 dw  \frac{  w^{h-1}}{\sqrt{ w + \sinh^2 { \ell\over 2} } } 
~_2F_1(h,h,1;1-w)\,,
\end{equation}
with $h \equiv { \Delta_\phi \over 2}$.  Note that $G_b$ is the bulk to boundary propagator for an operator of dimension $\Delta_\phi$, not to be confused with the $\chi$ field propagator  in \nref{ChiP}. The integral gives\footnote{ See 
  Gradshteyn and Ryzhik formula (7.512.9).}  
\bea
\langle\mathcal{O}\rangle &\propto &
\frac{\Gamma(h)\Gamma(1-h) } {  \sinh^2 { \ell \over 2 } }
~_3F_2\left(h,\half ,1-h;1,1;- { 1 \over \sinh^2 { \ell \over 2} }\right)
 \la{Finin} \eea
 If we now expand for large $\ell$, we can evaluate the hypergeometric function at zero and we get 
\begin{equation}
\langle\mathcal{O}\rangle
 \propto  \Gamma(h)\Gamma(1-h)\times
(\text{indep. of h})
\times e^{-\ell }
\approx \frac{1}{\sin(\pi h)}\times
(\text{indep. of h})
\times e^{-\ell}\,,
\end{equation}
This agrees with our previous $e^{-i\pi h}$ phase factor.  
The other terms in the sum over $n$ can be obtained by replacing $\ell \to n \ell$ in \nref{Finin} and give a similar $h$ dependence. The expected factor of $T^{\Delta_\phi}$ is obtained by rescaling the temperature back from $1/(2\pi)$ to the actual temperature. 

Therefore, we have checked that the BTZ result is reproducing the expectations we had in general, giving us a ``phase'' which is related to the time between the horizon and the BTZ singularity.  We have reproduced this only for the case of
$\Delta_{\chi}=1$. We also expect it to be true for more general values, but we have not explicitly performed the computation. 

Note that the answer has an exponential suppression $e^{-\ell} = 
 e^{ - \Delta_\chi \ell}$, as discussed in \cite{Kraus:2002iv}, since the one-point function must vanish for the black string case.

\section{Operator mixing} 
\la{OpMix}

In this appendix we discuss in more detail the operator mixing problem. 

Let us use the ``extrapolate'' dictionary to compute the one-point function \cite{Banks:1998dd}.  In this case, we are supposed to compute the expectation value of the bulk field, 
$\langle \phi(z')\rangle$, as a function of the bulk coordinate $z'$. We can then expand for small $z'$ and pick out the expectation value from the behavior 
\be 
\langle \phi(z')  \rangle \sim  \langle O \rangle (z')^\Delta\,,
\ee
where $\langle O \rangle$ is, by definition,  the coefficient of the $(z')^\Delta $ term. 

For the black brane case, the  expectation value of the bulk field is  
\be \la{TrueAn}
\langle \phi \rangle \sim \int { d z \over z^{d+1} } z^{ 2 d } G(z|z') \,,
\ee 
where $G(z|z')$ is the propagator.  In writing \nref{IntRe} we have assumed that $z'$ is very small and approximated the propagator in terms of the solution that is smooth at the horizon times a factor of $(z')^\Delta$ (which is correct for $z> z'$). In particular, we neglected the contribution from the region $z<z'$. This is valid if $\Delta < 2d$, when the integral \nref{IntRe} is convergent. 
If $\Delta> 2d$, the integral \nref{TrueAn} is still finite, however its leading behavior as $z'\to 0$ is obtained as 
\be \la{FinSa}
\langle \phi(z') \rangle ~\propto~ (z')^{d-\Delta} \int_{0}^{z'} { d z \over z^{d+1} } z^{ 2 d } z^\Delta  + (z')^\Delta \int_{z'}^{z_h} { d z \over z^{d+1} } z^{ 2 d }  z^{d-\Delta} ~~\propto~~  c_1 (z')^{2d} + c_2 (z')^\Delta \,,
\ee
where we indicated only the small $z,~z'$ behavior and $c_1, c_2$ are constants. The first term does not behave at all like an operator of dimension $\Delta$. Instead, it behaves like an operator of dimension $2d$. The idea is that this     term   is related to the expectation value of the square of the stress tensor and that the field $\phi$ contains both a contribution from the stress tensor, as well as a contribution from the primary field of dimension $\Delta$ that we are after. The latter is now contained in a subleading term. In writing \nref{IntRe} we have neglected the first term in \nref{FinSa} and approximated  $z'=0$ in the lower integration limit of the second term. The analytic continuation of the integral that we discussed is designed to pick out the piece going like $(z')^\Delta$, neglecting the terms related to the stress tensor.

\small
\bibliographystyle{ourbst}
 \bibliography{DraftOnePointFunctions.bib}
\end{document}